\documentclass[usenatbib]{mnras}
\usepackage[utf8]{inputenc}
\usepackage{aas_macros}
\usepackage{amsfonts}
\usepackage{amsmath}
\usepackage{amssymb}
\usepackage{graphicx}
\usepackage[dvipsnames]{xcolor}
\usepackage{hyperref}
\hypersetup{colorlinks=true,allcolors=teal}
\usepackage{microtype}

\newcommand{\Hi}{\textsc{Hi}}

\newcommand{\SHi}{\ensuremath{S_{\Hi}^{\rm (int)}}}

\newcommand{\p}{\ensuremath{\partial}}

\newcommand{\Mh}{\ensuremath{h^{-1}M_{\odot}}}
\newcommand{\Mhsq}{\ensuremath{h^{-2}M_{\odot}}}
\newcommand{\Mpch}{\ensuremath{h^{-1}{\rm Mpc}}}
\newcommand{\kpch}{\ensuremath{h^{-1}{\rm kpc}}}
\newcommand{\kms}{\ensuremath{{\rm \,km\,s}^{-1}}}

\newcommand{\avg}[1]{\ensuremath{\left\langle \,#1\, \right\rangle}}
\newcommand{\e}[1]{\ensuremath{{\rm e}^{#1}}}

\newcommand{\der}{\ensuremath{{\rm d}}}

\newcommand{\dir}{\ensuremath{\delta_{\rm D}}}

\newcommand{\eqn}[1]{equation~\eqref{#1}}
\newcommand{\eqns}[1]{equations~\eqref{#1}}
\newcommand{\ph}[1]{\phantom{#1}}

\newcommand{\be}{\begin{equation}}
\newcommand{\ee}{\end{equation}}
\newcommand{\Cal}[1]{\ensuremath{\mathcal{#1}}}

\title[HI velocity profiles in LCDM]{The distribution of HI velocity profiles in a $\Lambda$CDM universe} 
\author[Paranjape et al.]
{
{\parbox[t]{\textwidth}{
Aseem Paranjape$^{1}$\thanks{E-mail: aseem@iucaa.in},
R. Srianand$^{1}$\thanks{E-mail: anand@iucaa.in},
Tirthankar Roy Choudhury$^{2}$\thanks{E-mail: tirth@ncra.tifr.res.in} \&
Ravi K. Sheth$^{3,4}$\thanks{E-mail: shethrk@physics.upenn.edu}
}}
\\\, \\
 $^1$ Inter-University Centre for Astronomy \& Astrophysics,
      Ganeshkhind, Post Bag 4, Pune 411007, India\\
  $^2$ National Centre for Radio Astrophysics, TIFR, Post Bag 3, Ganeshkhind, Pune 411007, India\\
  $^3$ Center for Particle Cosmology, University of Pennsylvania, 209 S. 33rd St., Philadelphia, PA 19104, USA\\
 $^4$ The Abdus Salam International Center for Theoretical Physics, Strada Costiera, 11, Trieste 34151, Italy}

\begin{document}
\label{firstpage}
\pagerange{\pageref{firstpage}--\pageref{lastpage}}
\maketitle

\begin{abstract}
We model the distribution of the observed profiles  of 21 cm line emission from neutral hydrogen (\Hi) in central galaxies selected from a statistically representative mock catalog of the local Universe in the Lambda-cold dark matter framework. 
The distribution of these \Hi\ velocity profiles (specifically, their widths $W_{50}$) has been observationally constrained, but has not been systematically studied theoretically. 
Our model profiles derive from rotation curves of realistically baryonified haloes in an $N$-body simulation, including the quasi-adiabatic relaxation of the dark matter profile of each halo in response to its baryons. 
We study the predicted $W_{50}$ distribution  using a realistic pipeline applied to noisy profiles extracted from our luminosity-complete mock catalog with an ALFALFA-like survey geometry and redshift selection. 
Our default mock is in good agreement with observed ALFALFA results for $W_{50}\gtrsim700\kms$, being incomplete at lower widths due to the intrinsic threshold of $M_r\leq-19$. 
Variations around the default model show that the velocity width function at $W_{50}\gtrsim300\kms$ is most sensitive to a possible correlation between galaxy inclination and host concentration, followed by the physics of quasi-adiabatic relaxation. 
We also study the \emph{excess kurtosis} of noiseless velocity profiles, obtaining a distribution which tightly correlates with $W_{50}$, with a shape and scatter that depend on the properties of the turbulent \Hi\ disk.  
Our results open the door towards using the shapes of  \Hi\ velocity profiles as a novel statistical probe of the baryon-dark matter connection. 
\end{abstract}


\section{Introduction}
\label{sec:intro}
\noindent
The distribution of baryons in the Universe, particularly those locked up inside galaxies, is of fundamental interest for theories of structure formation. In the context of the Lambda-cold dark matter ($\Lambda$CDM) paradigm, a key goal is to robustly establish and theoretically interpret the details of the observed galaxy-dark matter connection. A host of observational probes is typically employed in this exercise, ranging from the distribution of masses and baryonic content of galaxy clusters \citep{vikhlinin+09b,vikhlinin+09a}, to the clustering of galaxies in the local \citep{zehavi+11,eBOSS20} and high-redshift Universe \citep{delatorre+11,marulli+13,laurent+17}, to the effects of gravitational lensing on galaxy shapes \citep{vikram+15,heymans+21}, all the way down to spatially resolved spectroscopy yielding information on the stellar content and inter-stellar medium of individual galaxies \citep{bundy+15} and (for spiral galaxies) their rotation curves \citep{pss96,mrdb01}.

Galaxy rotation curves in particular have a long history as probes of not only galactic structure and content \citep[e.g.,][]{abp87,sr01,gentile+04} but also the nature of gravity itself \citep[][]{bbs91,boac01,apr16,lms16b,mls16}. For relatively nearby (distance $\lesssim100\Mpch$) rotationally supported galaxies, rotation curves can be measured using either optical observations of the stellar content or radio-frequency observations of the cold gas content by exploiting the 21 cm line transition of neutral hydrogen (\Hi) \citep[][]{begeman89,boac01,boacs04,lms16b}. At larger distances ($z\sim0.1$), spatially resolved spectroscopy at radio frequencies becomes increasingly challenging due to the decreasing projected sizes of galaxies. Nevertheless, due to the  \emph{velocity} resolution of $\lesssim$ few \kms\ achieved by current radio telescopes, the rotation curves of \Hi-bearing galaxies can still be indirectly probed by observing the spatially integrated \Hi\ \emph{velocity profiles} -- i.e., the redshifted 21 cm flux as a function of observed frequency -- of individual objects. This quantity forms the key observable in large-volume surveys of \Hi-selected galaxies such as the \Hi\ Parkes  All Sky Survey \citep[HIPASS,][]{barnes+01,meyer+04}  or the Arecibo Legacy Fast ALFA (ALFALFA) survey \citep{giovanelli+05,giovanelli+07} and is the main focus of the present work. 
Ongoing and upcoming surveys of \Hi-bearing galaxies with the SKA precursors are expected to be wider and deeper than the present ones (e.g., WALLABY and DINGO using ASKAP, \citealp{2012MNRAS.426.3385D,2020Ap&SS.365..118K} and LADUMA using MeerKAT, \citealp{2012IAUS..284..496H}), which will extend the scope of studies like that presented in this paper.

There has been extensive work in the literature on the modelling of rotation curves in the $\Lambda$CDM framework, focused mainly on describing observed rotation curves by fitting them with static or dynamic mass models of the respective galaxy's baryonic and dark matter content \citep[see, e.g.,][]{abp87,gentile+04,bc04,granados+17,kckp20}. In recent work, some of us have explored an alternate route, using synthetic rotation curves -- produced as part of statistically representative mock galaxy catalogs -- to \emph{predict} the statistical properties of large samples of rotation curve data. The underlying mock catalogs are generated by populating gravity-only cosmological $N$-body simulations with galaxies, using an empirical halo occupation distribution (HOD) constrained by the observed galaxy abundances and luminosity-dependent clustering (see below).
The present work continues along these lines, focusing on self-consistently predicting the observed distribution of velocity profiles of massive \Hi-bearing galaxies in large surveys. 
The main motivation behind this exercise is the realisation that \Hi\ velocity profiles are, in principle, sensitive to a number of baryonic physics details due to their connection with the underlying rotation curve and the nature of the \Hi\ disk, as described in detail below. To our knowledge, this aspect of \Hi\ velocity profiles has not been systematically explored or exploited in the literature previously. The only works we are aware of are by \citet{papastergis+11} and \citet{moorman+14} who presented measurements of the  distribution of the velocity widths of \Hi-selected galaxies in the ALFALFA survey. As such, the distribution of shapes of \Hi\ velocity profiles is a hitherto unexplored probe of the baryon-dark matter connection at small scales.

With this in mind, in this work we explore the sensitivity of \Hi\ velocity profiles to various aspects of the baryon-dark matter physics, such as (i) scaling relations involving disk sizes, (ii) environmental effects, (iii) the physics of quasi-adiabatic relaxation of dark matter in the presence of baryons and (iv) the impact of baryonic physics involving the intrinsic dispersion of the  \Hi\ 21cm line in a galactic disk. As mentioned above, we perform this analysis using a realistic mock catalog of low-redshift ($z\lesssim0.1$) galaxies which is constrained to reproduce the abundances and clustering of optically selected galaxies in the Sloan Digital Sky Survey \citep[SDSS,][]{york+00},\footnote{\url{www.sdss.org}} and \Hi-selected galaxies in the ALFALFA survey. As part of our analysis, we perform an in-depth study of the extraction of velocity widths from our simulated velocity profiles in the presence of realistic noise, allowing us to compare with the published ALFALFA results from \citet{papastergis+11} and \citet{moorman+14}. Additionally, we emphasize the utility of beyond-width statistics such as excess kurtosis as a novel probe of baryonic physics in \Hi\ disks.

The paper is organised as follows. In section~\ref{sec:mock} we describe our mock catalogs and the procedure to `baryonify' the host halo of each \Hi-bearing central galaxy. In section~\ref{sec:HIvp}, we show how the rotation curve of such a galaxy can be used to model the \Hi\ profile it would present to a distant observer, discussing in detail the sensitivity of the model to different parameters and assessing its potential as a mass-modelling tool. We further discuss the extraction of the velocity width from a velocity profile in the presence of realistic noise, along with the subsequent estimate of the distribution of widths of an \Hi-selected sample. In section~\ref{sec:results}, we present the results of applying this procedure for obtaining the velocity width  function to our mock galaxy catalog, exploring a number of variations in sample selection and modelling choices around our default model, as mentioned above. In section~\ref{sec:excesskurtosis}, we move beyond the velocity width and propose the excess kurtosis of the velocity profile as a novel probe of the physics of turbulence in the \Hi\ disk. We summarise and conclude in section~\ref{sec:conclude}. The appendices present technical details related to some aspects of the analysis. Throughout, we assume a spatially flat $\Lambda$CDM background cosmology, with parameters $\{\Omega_{\rm m},\Omega_{\rm b},h,n_{\rm s},\sigma_8\}$ given by $\{$0.276, 0.045, 0.7, 0.961, 0.811$\}$, compatible with the 7-year results of the \emph{Wilkinson Microwave Anisotropy Probe} experiment \citep[WMAP7,][]{Komatsu2010}. We denote the base-10 (natural) logarithm as $\log$ ($\ln$).

\section{Mock galaxy catalog}
\label{sec:mock}
The mock galaxy catalog on which we build our analysis is constructed using the algorithm described by \citet[][hereafter, PCS21]{pcs21} and summarised below.

\subsection{Simulation and mock algorithm}
\label{subsec:sim}
\noindent
In this work, we rely on one realisation of the ${\rm L}300\_{\rm N}1024$ simulation box discussed by PCS21. The (gravity-only) simulation evolved $1024^3$ particles in a $(300\Mpch)^3$ cubic box with the code \textsc{gadget-2} \citep{springel:2005}\footnote{\url{http://www.mpa-garching.mpg.de/gadget/}}. Dark haloes were identified using the code \textsc{rockstar} \citep{behroozi13-rockstar}\footnote{\url{https://bitbucket.org/gfcstanford/rockstar}} and relaxed objects were retained, discarding substructure. Further details of the simulation can be found in \citet{pa20}. In the following, $m_{\rm vir}$ and $R_{\rm vir}$ will refer to the total halo mass and virial radius. We define $R_{\rm vir}\equiv R_{\rm 200c}$, the radius at which the enclosed halo-centric density becomes 200 times the critical density $\rho_{\rm crit}$ of the Universe, so that $m_{\rm vir} = (4\pi/3)R_{\rm vir}^3\times200\rho_{\rm crit}$.

Mock central and satellite galaxies were populated in these host haloes using the PCS21 algorithm to produce a luminosity-complete sample of galaxies with an $r$-band absolute magnitude threshold $M_r\leq-19$. This algorithm is based on the halo occupation distribution (HOD) model and optical-\Hi\ scaling relation calibrated by \citet{pcp18} and \citet{ppp19}, and additionally assigns each mock galaxy with realistic values of $g-r$ and $u-r$ colours and stellar mass $m_\ast$. Most importantly for the present work, approximately $60\%$ of these galaxies are also assigned values of neutral hydrogen (\Hi) mass $m_{\Hi}$ sampled from the optical-\Hi\ scaling relation. The HOD models underlying the algorithm are constrained by the observed abundances and clustering of optically selected galaxies in the SDSS and of \Hi-selected galaxies in the ALFALFA survey. The luminosity threshold of $M_r\leq-19$ leads to completeness limits of $10^{9.85}\Mhsq$ and $10^{9.7}\Mhsq$ in $m_\ast$ and $m_{\Hi}$, respectively. We refer the reader to PCS21 for various  tests and predictions of the algorithm.

\subsection{Baryonification and rotation curves}
\label{subsec:vrot}
\noindent
The host haloes of the central galaxies thus produced are `baryonified' by the PCS21 algorithm according to a modified version of the prescription of \citet[][hereafter, ST15]{st15} which we discuss next, focusing on galaxies containing \Hi.  
The host halo of each \Hi-bearing central galaxy is assigned spatial distributions of the following baryonic components:
\begin{itemize}
    \item A 2-dimensional axisymmetric \Hi\ disk (`\Hi') with scale length $h_{\Hi}$ (surface density $\Sigma_{\Hi}(r_\perp)\propto\e{-r_\perp/h_{\Hi}}$ in the disk plane), for centrals with an assigned $m_{\Hi}$ value. 
    The scale length $h_{\Hi}$ is assumed to follow the empirical scaling $h_{\Hi} \propto m_{\Hi}^{0.5}$ \citep[][see equation~8 of PCS21]{wang+16-HI}.
    The corresponding mass fraction is $f_{\Hi}=1.33\,m_{\Hi}/m_{\rm vir}$, with the prefactor accounting for Helium correction.
    \item A spherical distribution of stars in the central galaxy (`cgal') with half-light radius $R_{\rm hl}$ constrained by observations \citep{kravtsov13} and a mass fraction $f_{\rm cgal}=m_\ast/m_{\rm vir}$. The model currently does not include a separate stellar disk, which remains an interesting future extension. 
    \item Spherical distributions of gravitationally bound, hot ionized gas (`bgas') in hydrostatic equilibrium, and expelled gas (`egas') or the circum-galactic medium affected by feedback processes. The mass fraction $f_{\rm bgas}$ is extrapolated to low $m_{\rm vir}$ from the relation calibrated by ST15 from X-ray cluster observations; $f_{\rm bgas}\lesssim0.01$ for typical \Hi-bearing centrals. The mass fraction $f_{\rm egas}$ is set by baryonic mass conservation (see PCS21 for details).
\end{itemize}
Finally, the presence of these baryonic components is assumed to backreact on the dark matter profile according to the prescription of ST15 (see appendix~A of PCS21), leading to a quasi-adiabatic relaxation, approximately conserving angular momentum, which tends to contract the dark matter in the inner halo and slightly expand it the halo outskirts, on average \citep[see, e.g., fig.~1 of][]{ps21}. The physics of this relaxation is parametrised by a quantity $q_{\rm rdm}$ (e.g., equation~A1 of PCS21), such that $q_{\rm rdm}=0$ corresponds to no baryonic backreaction and $q_{\rm rdm}=1$ to perfect conservation of angular momentum. The default value adopted in the PCS21 mocks and used below is $q_{\rm rdm}=0.68$, which was suggested by ST15 based on the hydrodynamical CDM simulation results of \citet{teyssier+11}.

We refer the reader to section~3.2 of PCS21 for details of the numerical implementation of this scheme, as well as all the underlying scalings of baryonic mass fractions and galaxy sizes with halo properties.  
Baryonification schemes of this type have been shown to successfully reproduce the small-scale spatial correlation statistics of cosmological hydrodynamical simulations \citep[e.g.,][]{chisari+18,arico+20b}.

The spatial distributions of baryons and dark matter produced by the scheme above allow for a calculation of the rotation curve of each mock central galaxy. For \Hi-bearing galaxies, we focus on the mid-plane of the thin exponential \Hi\ disk, which gives a circular velocity contribution $v_{\Hi}(r)$ satisfying  
\be
v_{\Hi}^2(r) = \frac{2f_{\Hi}V_{\rm vir}^2}{\left(h_{\Hi}/R_{\rm vir}\right)}\,y^2\left[I_0(y)K_0(y) - I_1(y)K_1(y)\right]\,,
\label{eq:expdisk-vrot}
\ee
where $y\equiv r/(2h_{\Hi})$, $V_{\rm vir} = \sqrt{Gm_{\rm vir}/R_{\rm vir}}$ is the virial velocity and $I_n(y)$ and $K_n(y)$  are  modified Bessel functions of the first and second kind, respectively.
The rotation curve $v_{\rm rot}(r)$ for each mock galaxy is calculated using equation~(11) of PCS21, which can be rewritten as
\begin{align}
v_{\rm rot}^2(r) &= v_{\Hi}^2(r) + \sum_{\alpha} \frac{Gm_{\alpha}(<r)}{r} +  \frac{Gm_{\rm rdm}(<r)}{r}\,,
\label{eq:vrot-def}
\end{align}
where the sum runs over $\alpha\in\{{\rm bgas},{\rm cgal},{\rm egas}\}$, $m_{\alpha}(<r)$ is the mass of component $\alpha$ enclosed in radius $r$ and $m_{\rm rdm}(<r)$ is the corresponding mass of the quasi-adiabatically relaxed dark matter component. 
The rotation curves produced by the default baryonification model adopted by PCS21 have been shown to be in very good agreement with the median and scatter of the radial acceleration relation of low-redshift galaxies \citep{ps21}.

\section{Modelling HI velocity profiles}
\label{sec:HIvp}
\noindent
A mock rotation curve, along with an assignment of an `observed' redshift (see appendix~\ref{app:rsd}) and inclination angle to the galaxy, can  be used to predict the observed velocity profile of the \Hi\ 21 cm emission line in a survey such as ALFALFA. In this section, we describe our methodology to predict the \Hi\ velocity profile $S_{\Hi}(v)$ for each central galaxy, followed by an assessment of its potential as a mass-modelling tool, and a description of our technique for extracting the velocity width $W_{50}$ in realistic observational samples.

\subsection{From rotation curves to velocity profiles}
\label{subsec:vrot-Hivp}
\noindent
The rotation curve of each \Hi-bearing galaxy can be converted into the observable $S_{\Hi}(v)$ essentially using geometrical considerations and accounting for the Doppler-shifting of line emission from a differentially rotating system \citep{gordon71,roberts78}. We consider a thin \Hi\ disk as described in section~\ref{subsec:vrot}, inclined at an angle $i$ relative to the observer's line of sight (such that $i=0^\circ$ for a face-on disk). We assume the optically thin regime, which is a good approximation for all but nearly edge-on disks. Finally, we assume that the observed \Hi\ 21 cm line has an intrinsic Gaussian velocity distribution $p(v)$ \citep{sbr94} with width $\sigma_v\lesssim10\,\kms$ arising from turbulent motions in the disk \citep{sb99}. 

\begin{figure*}
\centering
\includegraphics[width=0.43\textwidth]{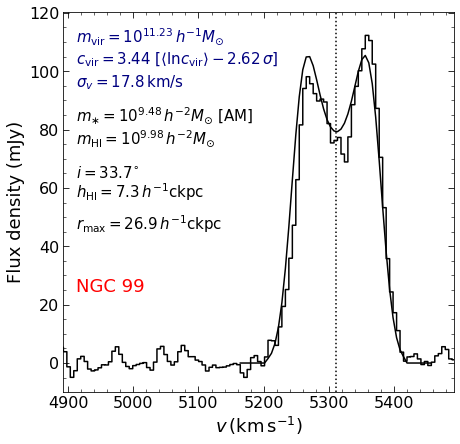}
\includegraphics[width=0.45\textwidth]{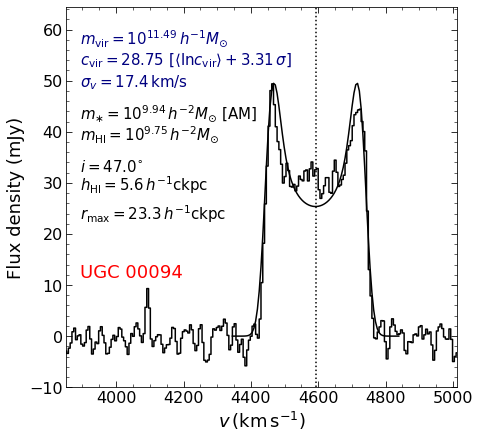}
\caption{{\bf Velocity profile model for two observed galaxies}. The stepped black line in the \emph{left (right) panel} shows the observed profile of NGC 99 (UGC 00094) from the ALFALFA  source catalog \citep{haynes+18}, while the smooth black curve shows our best fit model. 
For each galaxy, the values of inclination $i$, $m_{\Hi}$, $h_{\Hi}$ and $r_{\rm max}$ were fixed as described in the text.
The remaining parameters $m_{\rm vir}$, $c_{\rm vir}$ and  $\sigma_v$ were varied in a least squares calculation to obtain the best-fit values (marked in blue), with the stellar mass $m_\ast$ set by abundance matching (AM) as described in the text.
The profiles are centered at the systemic velocities reported by \citet{haynes+18}: $cz=5312\,(4592)\,\kms$ for NGC 99 (UGC 00094) shown as the vertical dotted line in each panel.}
\label{fig:HIvp-obs}
\end{figure*}

The observed flux density $S_{\Hi}(v)$ in a velocity channel $(v,v+\der v)$ then satisfies \citep{gordon71,sbr94}
\begin{align}
S_{\Hi}(v) &\propto \int\der v^\prime\,p(v-v^\prime)\int_0^\pi\der\theta\int\der r_\perp\,2\,r_\perp\Sigma_{\Hi}(r_\perp)\,B(r_\perp,\theta) \notag\\
&\ph{\int\der v^\prime\,p(v-v^\prime)\int_0^\pi\der\theta\int}
\times\dir\left(v^\prime - v_{\rm rot}(r_\perp)\sin i\cos\theta\right) \notag\\
&\approx \int_0^{r_{\rm max}}\der r_\perp\,2\,r_\perp\Sigma_{\Hi}(r_\perp)\notag\\
&\ph{\int_0^{r_{\rm max}}\der r_\perp}
\times\int_0^\pi\der\theta\,p\left(v - v_{\rm rot}(r_\perp)\sin i\cos\theta\right)\,.
\label{eq:SHI}
\end{align}
Here $\dir(x)$ is the Dirac delta distribution, $r_\perp$ and $\theta$ are the radial distance and azimuthal angle, respectively, in the disk plane and $B(r_\perp,\theta)$ is the telescope beam response converted to the disk reference frame. In the second line, we approximated the beam response as a simple tophat in $r_\perp$. Throughout, we will assume the relation
\be
r_{\rm max} = (\theta_{\rm beam}/2)\times D_{\rm A}(z)\,, 
\label{eq:rmax-def}
\ee
where $D_A(z)$ is the angular diameter distance to redshift $z$ and $\theta_{\rm beam}$ is the instrument beam width in radians.

The normalisation of $S_{\Hi}(v)$ is fixed by relating its integral to the \Hi\ mass $m_{\Hi}$ and luminosity distance $D_{\rm L}$ of the galaxy \citep{roberts75,gh88}
\be
m_{\Hi} = 2.356\times10^5\,\Mhsq\left(\frac{D_{\rm L}}{\Mpch}\right)^2 \, \int\frac{\der v}{\kms}\frac{S_{\Hi}(v)}{\rm Jy}\,.
\label{eq:mHI-SHI}
\ee
We emphasize that $S_{\Hi}(v)$ is sensitive to the entire matter content of the galaxy's host halo (stars, neutral gas, ionised gas, and their effect on the dark matter), not just the \Hi\ disk, through its dependence on the rotation curve $v_{\rm rot}(r)$ in \eqn{eq:SHI}. Our analysis self-consistently produces the velocity profile and rotation curve for a given baryonic composition of the host halo without, e.g., treating the rotation curve independently of the \Hi\ disk. In principle, the model can be made more complex by including the effects of (i) holes and warps in the \Hi\ surface density $\Sigma_{\Hi}$, e.g., by separately modelling a stellar and gas disk, (ii) high velocity clouds (HVCs) modelled by changing the intrinsic velocity distribution $p(v)$ \citep{sbr94} or (iii) a more realistic beam profile $B(r_\perp,\theta)$ \citep{gordon71}. We will ignore the first two for simplicity, while the third is unlikely to be relevant for large beams which do not resolve individual galaxies.

\subsection{Examples: parameter inference and sensitivity}
\label{subsec:examples}
\noindent
In this section, we compare the results of numerically integrating the double integral in \eqn{eq:SHI} with two example \Hi\ velocity profiles of real galaxies, by adjusting some of the model parameters.
This allows us to assess the potential of our model as a mass-modelling parameter inference tool, and also explore its sensitivity to various parameters. Although not our primary aim in this work, this exercise will inform our subsequent exploration of the statistical distributions of velocity profile properties.

\subsubsection{Modelling NGC 99 and UGC 00094}
\label{subsubsec:ngc99ugc00094}
We consider two galaxies, NGC 99 and UGC 00094, whose velocity profiles we obtain from the ALFALFA  source catalog presented by \citet{haynes+18}. \citep[NGC 99 was also modelled using early Arecibo observations by][see their fig.~2]{sbr94}. In  each case, we fix the values of $m_{\Hi}$ and $D_{\rm L}$ using, respectively, the integrated flux from the observed profile and the systemic velocity reported by \citet{haynes+18}. The value of the disk scale length $h_{\Hi}$ is then fixed using the empirical scaling relation mentioned in section~\ref{subsec:vrot}. We use the inclination reported by \citet[][their table 1]{sanchez+12} and \citet[][their table A.1]{tf14} for NGC 99 and UGC 00094, respectively, and use the Arecibo beam width of $\theta_{\rm beam}\simeq3.5^\prime$ to set $r_{\rm max}$ using \eqn{eq:rmax-def}. We then vary the values of the remaining parameters, namely halo mass $m_{\rm vir}$, halo concentration $c_{\rm vir}$ and intrinsic dispersion $\sigma_v$, using $m_{\rm vir}$ to fix the stellar mass $m_\ast$ using the abundance matching (AM) prescription of \citet{behroozi+13-SHAM}, with recalibrated parameter values from \citet{kvm18}. The values of $m_{\rm vir}$ and $c_{\rm vir}$ also fix other baryonification variables such as the stellar bulge size and the mass fractions and profiles of ionised and expelled gas (see section~\ref{subsec:vrot}). We hold the value of the relaxation parameter fixed at the default $q_{\rm rdm}=0.68$ in this exercise. The \emph{left (right) panel} of Fig.~\ref{fig:HIvp-obs} shows the observed and best-fit profile of NGC 99 (UGC 00094), along with the values of various parameters.

For NGC 99 (UGC 00094) the best-fit $m_{\rm vir}$ leads to a cold gas fraction $f_{\Hi}\simeq0.106 \,(0.035)$ and a gas-to-stellar mass ratio $f_{\Hi}/f_{\rm cgal}\simeq4.3\, (0.9)$, implying that NGC 99 is a relatively gas-rich system compared to UGC 00094. The log-concentration for NGC 99 (UGC 00094) is $\sim2.6\sigma$ lower ($\sim3.3\sigma$ higher) than the median value for each halo mass \citep[calculated using the calibration of][]{dk15}. Although these values represent statistically rare fluctuations relative to the $\Lambda$CDM expectation, we note that $c_{\rm vir}$ is strongly degenerate with $m_{\rm vir}$ when both are left free as in our case. Almost equally acceptable fits can also be achieved in each case with more reasonable $c_{\rm vir}$ values, by adjusting $m_{\rm vir}$. This degeneracy is difficult to break with \Hi\ velocity profiles alone.\footnote{For completeness, we report that using a simple NFW density profile and rotation curve without any baryonic components leads to completely unrealistic solutions; e.g., the inferred $c_{\rm vir}$ is more than $6\sigma$ away from the median relation.}

The best-fit value of the intrinsic dispersion $\sigma_v$ of $\sim17$-$18\kms$ in each example is substantially higher than the typical values of $6$-$10\kms$ reported for individual systems using spatially resolved spectroscopy \citep[e.g.,][]{sb99,stilp+13}. This could be due to unmodelled HVCs along the line-of-sight which can broaden the spatially integrated profile, especially affecting its tails \citep[see, e.g., the discusion in][who model NGC 99 and other galaxies including HVCs]{sbr94}. We have found that $\sigma_v$ is also degenerate with $m_{\rm vir}$ and $c_{\rm vir}$, so that improving the modelling of HVCs would also, in general, affect their inferred values. For example, fixing $\sigma_v=14\kms$ for NGC 99, as suggested by \citet{sbr94}, leads to best-fit values of $m_{\rm vir}=10^{11.03}\Mh$ and a log-concentration $0.33\sigma$ \emph{above} the median.

\begin{figure*}
\centering
\includegraphics[width=0.475\textwidth,trim=6 0 4 0,clip]{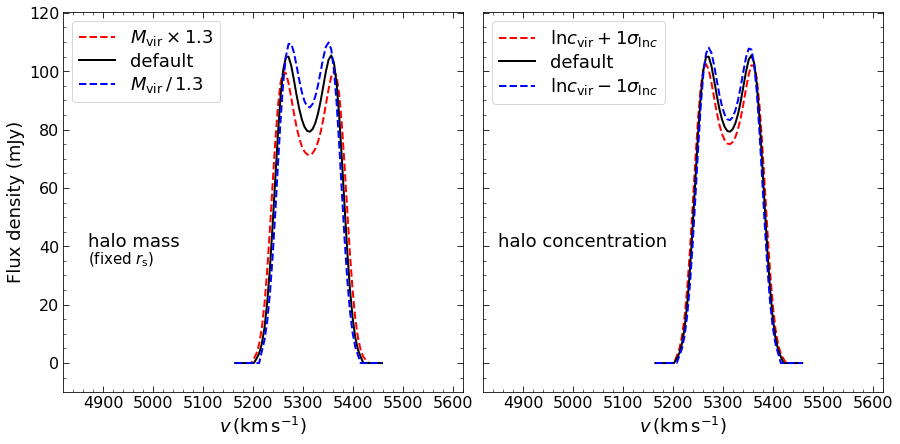}
\includegraphics[width=0.475\textwidth,trim=6 0 4 0,clip]{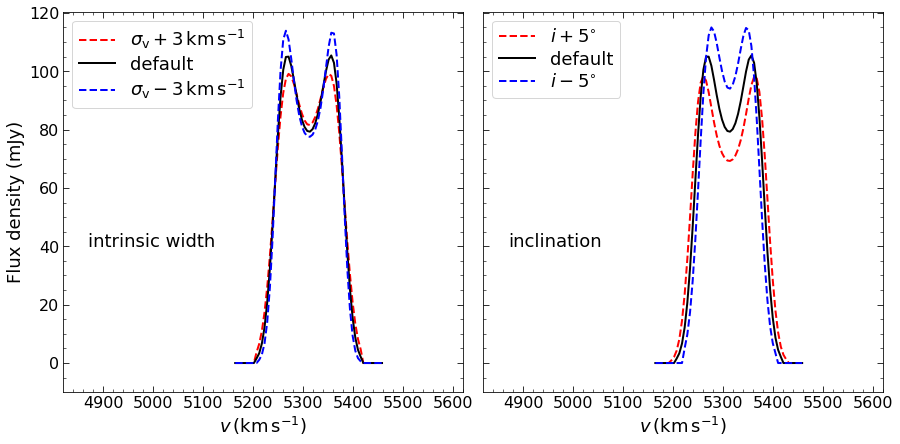}
\includegraphics[width=0.475\textwidth,trim=6 0 4 0,clip]{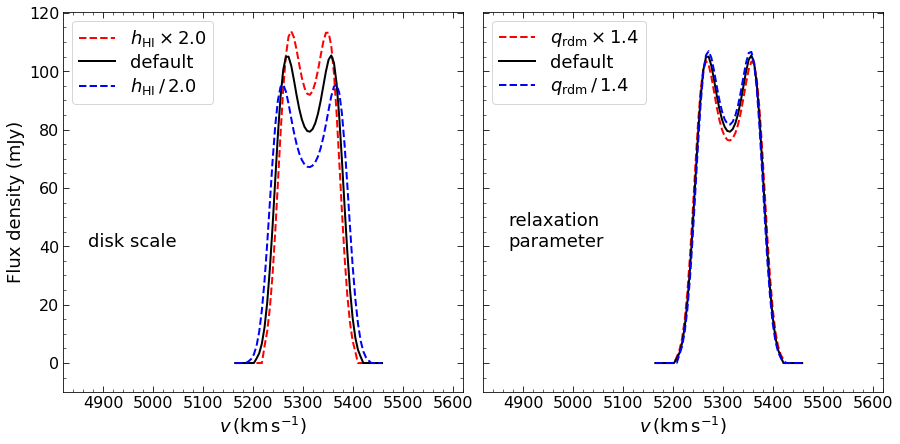}
\includegraphics[width=0.475\textwidth,trim=6 0 4 0,clip]{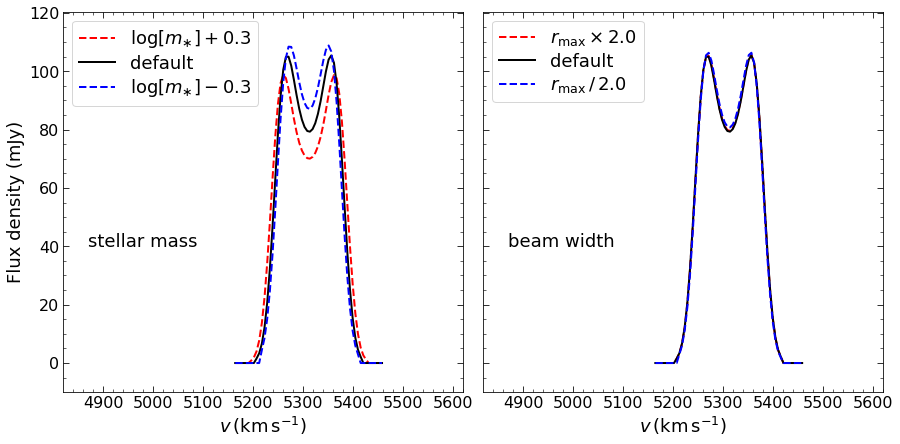}
\caption{{\bf Parameter sensitivity of velocity profile.} Each panel shows the result of varying one parameter at a time around the default model of NGC 99 from the left panel of Fig.~\ref{fig:HIvp-obs}, shown as the solid black curve in each panel. Upward (downward) variations of each parameter are shown as the red (blue) dashed curve in each panel.}
\label{fig:HIvp-NGC99-vary}
\end{figure*}

\begin{figure*}
\centering
\includegraphics[width=0.475\textwidth,trim=6 0 4 0,clip]{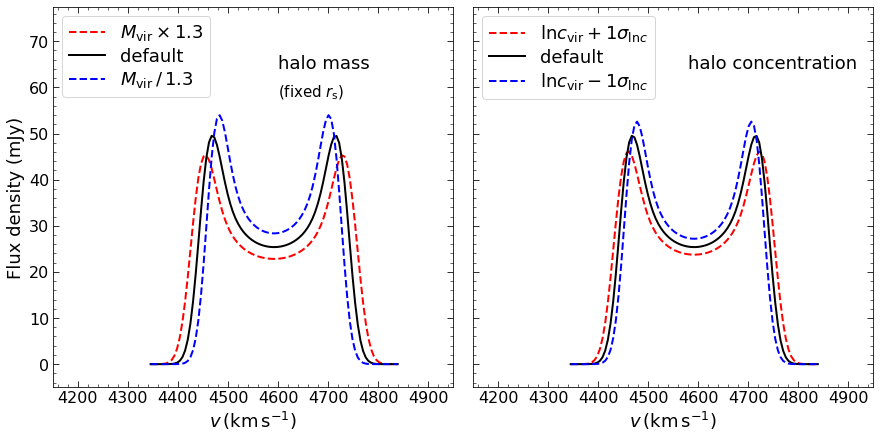}
\includegraphics[width=0.475\textwidth,trim=6 0 4 0,clip]{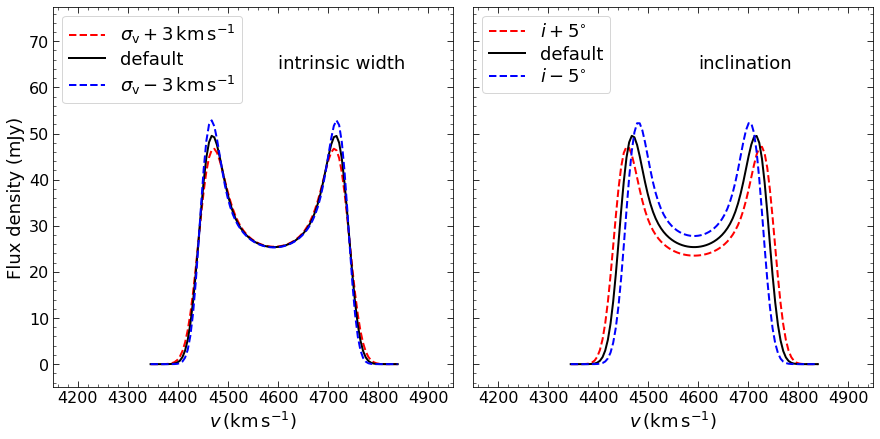}
\includegraphics[width=0.475\textwidth,trim=6 0 4 0,clip]{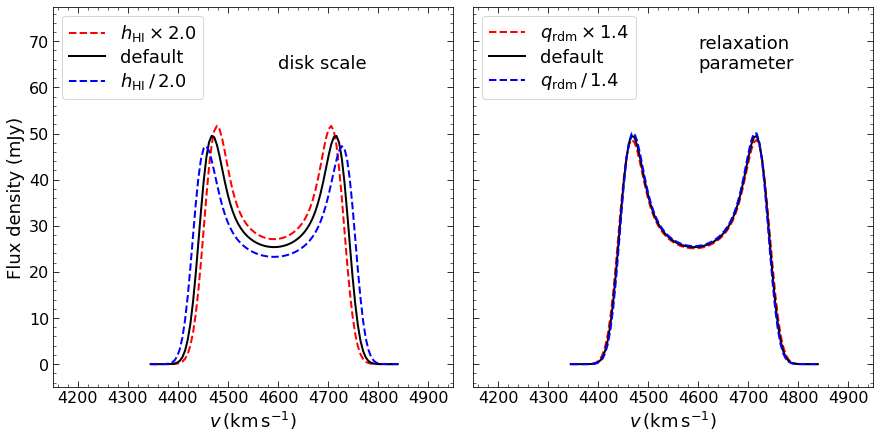}
\includegraphics[width=0.475\textwidth,trim=6 0 4 0,clip]{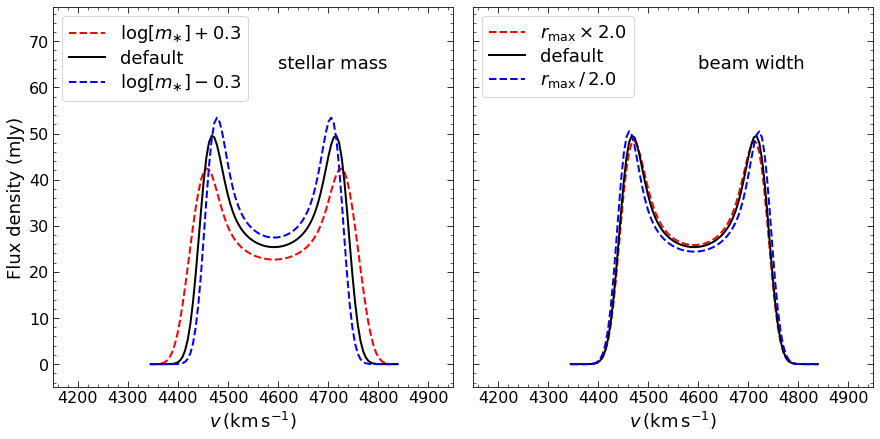}
\caption{Same as Fig.~\ref{fig:HIvp-NGC99-vary} but using the best fit model for UGC 00094 from the right panel of Fig.~\ref{fig:HIvp-obs} as the default.}
\label{fig:HIvp-UGC00094-vary}
\end{figure*}

\subsubsection{Sensitivity to parameter variations}
\label{subsubsec:varyparams}
To better understand some of the degeneracies discussed above, we next explore the effects on $S_{\Hi}(v)$ of varying each parameter individually, using our best-fit models for NGC 99 and UGC 00094 as the defaults. We do not attempt to model the obvious asymmetry between the two horns of each observed profile in Fig.~\ref{fig:HIvp-obs}, commenting on this aspect at the end of the section. We display results for NGC 99 (UGC 00094) in Fig.~\ref{fig:HIvp-NGC99-vary} (Fig.~\ref{fig:HIvp-UGC00094-vary}) for variations of $m_{\rm vir}$, $c_{\rm vir}$, $\sigma_v$, inclination $i$, $h_{\Hi}$, $m_\ast$, $r_{\rm max}$ and the relaxation parameter $q_{\rm rdm}$, with the default case repeated as the black curve in each panel. For the chosen default parameter values, the model is visibly most sensitive to $h_{\Hi}$ and $i$, followed by $m_{\rm vir}$, $c_{\rm vir}$ and $m_\ast$, while being less sensitive to $\sigma_v$, $q_{\rm rdm}$ and $r_{\rm max}$ (the last is understandable due to the large width of the Arecibo beam in comparison to the sizes of our chosen galaxies).

Most of these trends can be understood by inspecting \eqn{eq:SHI}. For an exponential surface density $\Sigma_{\Hi}\propto\e{-r_\perp/h_{\Hi}}$, the locations of the two horns of the velocity profile are determined roughly by the combination $v_{\rm rot}(h_{\Hi})\sin i$. For example, increasing (decreasing) the inclination will cause the two horns to go further apart (come closer), making the profile broader (narrower) while keeping its integral fixed \citep{gordon71,sbr94}. This is exactly the trend seen in the \emph{upper right-most panels} of Figs.~\ref{fig:HIvp-NGC99-vary} and~\ref{fig:HIvp-UGC00094-vary} (see also Fig.~\ref{figapp:tempmatch}). Since the effect of $v_{\rm rot}(h_{\Hi})$ is identical to that of $\sin i$, any variation that increases or decreases $v_{\rm rot}(h_{\Hi})$ can be understood in the same manner. This is clearly the case for $m_{\rm vir}$ at fixed halo scale radius $r_{\rm s}=R_{\rm vir}/c_{\rm vir}$ \emph{(upper left-most panels)}: changing $m_{\rm vir}$ primarily scales the overall amplitude of $v_{\rm rot}$ by changing $V_{\rm vir}\propto m_{\rm vir}^{1/3}$, apart from other effects due to changes in the various baryonic fractions. Increasing (decreasing) $m_{\rm vir}$ thus has a qualitatively similar effect to increasing (decreasing) $\sin i$. Similar reasoning also explains the trend seen with halo concentration $c_{\rm vir}$ at fixed $m_{\rm vir}$: high-concentration haloes tend to have higher  peak rotation curve values, and hence higher $v_{\rm rot}(h_{\Hi})$, as compared to low-concentration haloes of the same mass \citep[e.g.,][]{nfw96}, so that variations in $c_{\rm vir}$ are also qualitatively similar to those in $\sin i$. 

Variations in stellar mass $m_\ast$ (\emph{lower middle-right panels} of Figs.~\ref{fig:HIvp-NGC99-vary} and~\ref{fig:HIvp-UGC00094-vary}) behave very similarly to those in $m_{\rm vir}$ and $c_{\rm vir}$. Increasing (decreasing) $m_\ast$ affects the rotation curve \eqref{eq:vrot-def} in two ways: (i) it increases (decreases) the contribution of the stellar profile $m_{\rm cgal}(<r)$ in the inner halo and, consequently, (ii) it leads to a stronger (weaker) contraction of the dark matter profile. Both effects conspire to make the halo more (less) centrally concentrated, thus explaining the trend. (Similar results would be true if we simultaneously varied $m_{\Hi}$ and $D_{\rm L}$ keeping $m_{\Hi}/D_{\rm L}^2$ fixed.) And, as expected for the large Arecibo beam, the beam width variable $r_{\rm max}$ has a relatively minor effect, being more prominent for UGC 00094 which is the closer of the two systems.

The remaining three variables explored in Figs.~\ref{fig:HIvp-NGC99-vary} and~\ref{fig:HIvp-UGC00094-vary}, namely $\sigma_{v}$, $h_{\Hi}$ and $q_{\rm rdm}$, behave somewhat  differently than the others. The intrinsic dispersion
$\sigma_v$ affects the width of the distribution $p(v)$ in \eqn{eq:SHI} without changing its mean, so that increasing (decreasing) $\sigma_v$  makes each horn broader (narrower) without changing its position, as is clearly seen in the \emph{upper middle-right panels} of each figure. We will return to this effect in section~\ref{subsubsec:results-sigvSigHI}. The disk size $h_{\Hi}$ affects not only the location at which the rotation curve is effectively sampled due to the exponential surface density (as implied by our writing the combination $v_{\rm rot}(h_{\Hi})\sin i$ above), but also the shape of the rotation curve itself. For the rotation curves of \Hi-bearing centrals in our luminosity-complete mock catalog (section~\ref{sec:mock}), we find that increasing $h_{\Hi}$ for each galaxy while keeping all its other variables fixed tends to \emph{decrease} $v_{\rm rot}(h_{\Hi})$ on average, and vice-versa. This is consistent with the behaviour seen in the \emph{lower left-most panels}, where increasing (decreasing) $h_{\Hi}$ has an effect similar to decreasing (increasing) $\sin i$. 

The relaxation parameter $q_{\rm rdm}$ controls the amount of contraction or expansion of the dark matter profile due to the baryonic components. This is a novel aspect of our model which has been generally ignored in the mass-modelling literature. For the chosen default profiles, varying $q_{\rm rdm}$ has a weak effect \emph{(lower middle-left panels)}, being more noticeable for NGC 99 in Fig.~\ref{fig:HIvp-NGC99-vary}. The weakness of the effect follows from the fact that quasi-adiabatic relaxation largely affects the inner halo, while the double-horn structure of the \Hi\ profile is more sensitive to the peak or flat part of $v_{\rm rot}$. The trends seen are also sensible: a larger $q_{\rm rdm}$ leads to a stronger contraction of the dark matter profile of each halo, making it more centrally concentrated, so that the effect of $q_{\rm rdm}$ is qualitatively similar to that of  $c_{\rm vir}$ (e.g., compare the \emph{lower} and \emph{upper middle-left panels} of Fig.~\ref{fig:HIvp-NGC99-vary}).

\begin{figure*}
\centering      
\includegraphics[width=0.95\textwidth]{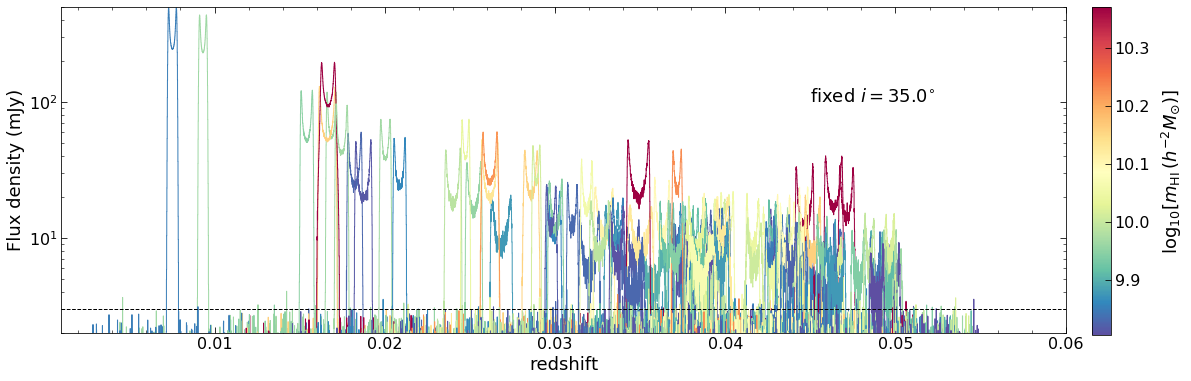}
\includegraphics[width=0.95\textwidth]{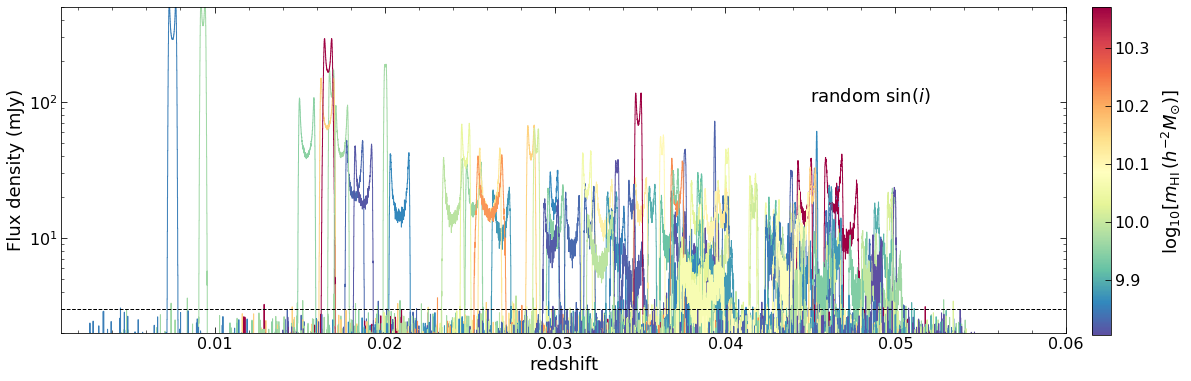}
\caption{{\bf Mock line profiles} of \Hi\ disks of 100 galaxies randomly chosen from a sample selected to have $M_r\leq-19$, $m_{\Hi}\geq10^{9.8}\Mhsq$ and $z\leq0.05$ in a $300\Mpch$ box with $2\,\kms$ channel widths. The profiles were generated using our default model as described in section~\ref{sec:HIvp}, with redshifts assigned as described in appendix~\ref{app:rsd} for arbitrary lines of sight and a Gaussian noise of $1\,{\rm mJy}$ per velocity channel added. The \emph{upper panel} shows results for galaxies observed with a fixed inclination angle of $i=35^\circ$, while the \emph{lower panel} shows the same galaxies with randomised inclinations ($\sin i$ uniformly sampled in the range $[0,1)$). Each curve is coloured by the value of \Hi\ mass $m_{\Hi}$ as indicated by the colour bar. With fixed inclination, we clearly see the overall decrease in amplitude due to increasing distance, with only a few high-mass objects jutting out over the envelope. With randomised inclinations, low-mass objects at higher redshift can also have high amplitudes. The horizontal dashed line indicates the $3\sigma$ noise level.}
\label{fig:mock-lineprofiles}
\end{figure*}

The reason NGC 99 shows a more prominent effect than UGC 00094 is more subtle, however. The effects of quasi-adiabatic relaxation on the rotation curve at the mass scales of our interest depend on the combination of $m_{\rm vir}$, $c_{\rm vir}$, $m_\ast$ and $m_{\Hi}$, along with the spatial extents of the stars and cold gas \citep[see, e.g., fig.~1 of][]{ps21}. To try and disentangle these effects, we varied the values of $m_\ast$ and $m_{\Hi}$ independently for each of these examples, producing three sets of curves in addition to  those shown in Figs.~\ref{fig:HIvp-NGC99-vary} and~\ref{fig:HIvp-UGC00094-vary}: one in which $m_{\Hi}$ is increased by 0.3 dex, one in which $m_\ast$ is increased by 0.3 dex and one in which both are increased by this amount while keeping their ratio fixed. In each case, we calculated the ratio of the profile widths obtained using the higher and lower values of $q_{\rm rdm}$. For both NGC 99 and UGC 00094, this ratio of widths is most sensitive to changes in $m_\ast$ at fixed $m_{\Hi}$, while the other two variations produced almost no effect. However, while the ratio rises from $\sim1.035$ to $\sim1.06$ for UGC 00094 when $m_\ast$ is increased, it \emph{decreases} from $\sim1.08$ to $\sim1.035$ for NGC 99.  That is to say, while our model for UGC 00094 becomes more sensitive to $q_{\rm rdm}$ when $m_\ast$ is increased, the opposite is true for NGC 99. This makes it interesting to ask how changes in $q_{\rm rdm}$ would affect the overall distribution of, say, profile widths for a statistically representative sample; we explore this later in section~\ref{subsubsec:results-relax}.\footnote{We have also checked that the effect of modifying the bound gas fraction scaling $f_{\rm bgas}(m_{\rm vir})$ is negligible, while only very large ($\gtrsim$ factor 2) variations in the stellar bulge size $R_{\rm hl}$ lead to appreciable changes in the velocity profiles of both NGC 99 and UGC 00094. We will therefore not discuss these two parameters further.}

As mentioned previously, our model ignores the asymmetry of the observed profiles. This could easily bias the inferred values of $m_{\rm vir}$ and $c_{\rm vir}$ due to their degeneracy. Asymmetry in observed profiles could arise due to several reasons, from effects such as beam mis-centering for relatively nearby or large galaxies, to physical effects on the galaxy's morphology caused by interactions between the stellar and \Hi\ disk or with the environment, particularly in dense regions \citep[see, e.g.,][]{bbge19,watts+20}. This would require making our disk model substantially more complex, with the inclusion of several new parameters. The lack of inherent asymmetry, and the fact that we do not model a stellar disk, also prevents us from testing the AM assumption by independently varying $m_\ast$: doing so leads to runaway behaviour, with extremely strong degeneracies appearing between $m_\ast$ and $m_{\rm vir}$ as expected from Figs.~\ref{fig:HIvp-NGC99-vary} and~\ref{fig:HIvp-UGC00094-vary}. Similarly, opening up the inclination angle as a free variable also leads to runaway behavior, indicating that knowledge of $\sin i$ for the \Hi\ disk is a minimum requirement if our model is to be used for parameter inference. To conclude this discussion, we note that our model produces reasonably realistic descriptions of symmetric profiles, while the modelling of asymmetries is currently challenging. 

\begin{figure*}
\centering
\includegraphics[width=0.95\textwidth]{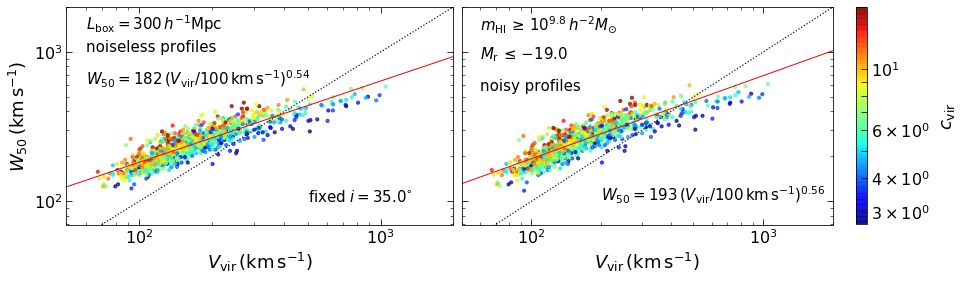}
\caption{{\bf Velocity width (fixed inclination).} $W_{50}$ measured from mock \Hi\ line profiles of 1000 galaxies randomly chosen from a sample selected to have $M_r\leq-19$, $m_{\Hi}\geq10^{9.8}\Mhsq$ and $z\leq0.05$ in a $300\Mpch$ box, observed with a \emph{fixed} inclination angle $i=35^\circ$ (see also Fig.~\ref{fig:mock-lineprofiles}). Each galaxy is shown as a marker in the plane of $W_{50}$ and halo virial velocity $V_{\rm vir}=\sqrt{Gm_{\rm vir}/R_{\rm vir}}$, coloured by the value of halo concentration $c_{\rm vir}$. The \emph{left (right) panel} shows results for noiseless (noisy) profiles, with noise corresponding to $1\,{\rm mJy}$ per $2\,\kms$ channel. There is evidently a tight relation between $\log W_{50}$ and $\log V_{\rm vir}$, quantified by the linear regression shown as the red solid line in each panel with parameters indicated in the labels. The dotted line in each panel shows the one-to-one relation for comparison. The effect of noise is clearly minimal, and the scatter around the mean relation is correlated with $c_{\rm vir}$: high-concentration haloes at fixed $V_{\rm vir}$ have larger $W_{50}$. }
\label{fig:mock-W50-fixedincl}
\end{figure*}

\begin{figure*}
\centering
\includegraphics[width=0.95\textwidth]{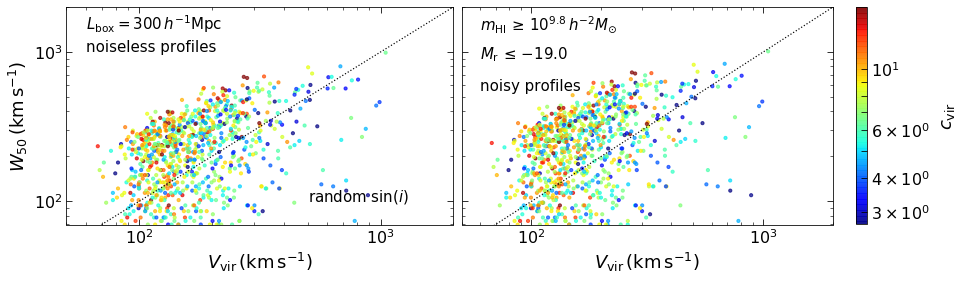}
\caption{{\bf Velocity width (random inclination).} Same as Fig.~\ref{fig:mock-W50-fixedincl}, for the same galaxies, now observed with randomised inclination angles. The scatter between $W_{50}$ and $V_{\rm vir}$ is now much broader, and the correlation between $W_{50}$ and halo concentration at fixed $V_{\rm vir}$ is also visibly weaker.}
\label{fig:mock-W50}
\end{figure*}

This machinery can be used to generate `observed' velocity profiles for our mock galaxies (in which all the parameters are known) by placing them in redshift space relative to an observer sitting at the center of one face of the simulation box. Appendix~\ref{app:rsd} describes our procedure to move galaxies into redshift space and assign them an observed redshift. Fig.~\ref{fig:mock-lineprofiles} shows a sample of noisy velocity profiles of \Hi-selected galaxies in our default mock catalog, with the \emph{upper panel} showing galaxies observed with a fixed inclination of $i=35^\circ$ and the \emph{lower panel} showing the same galaxies observed with random inclination angles. For simplicity, we set $\sigma_v=10\,\kms$ for all the objects. The 100 galaxies shown were randomly selected from a sample satisfying $M_r\leq-19$ and  $m_{\Hi}\geq10^{9.8}\Mhsq$. The profiles are coloured by the value of $m_{\Hi}$, and we see in the upper panel that only high-mass objects jut out over the envelope of decreasing amplitude as a function of distance. With randomised inclinations, on the other hand, low-mass objects can also be detected with high significance depending on how close to face-on they are viewed. For this example, we used $2\,\kms$ velocity channels with $1\,{\rm mJy}$ Gaussian noise added per channel, similar to the Arecibo observations used by \citet{sbr94}.  Below, we will discuss in detail the effects of noise in a realistic survey.

\subsection{Velocity widths from velocity profiles}
\label{subsec:W50-basicprops}
\noindent
For each observed mock profile, we estimate the velocity width $W_{50}$ 
using a modified version of the template-matching algorithm described by \citet{saintonge07}. This technique, which we describe in appendix~\ref{app:templatematching}, will also be used later when discussing realistic surveys. For the present exercise, we do not smooth the data and also do not place any restriction on signal-to-noise when selecting galaxies. Fig.~\ref{fig:mock-W50-fixedincl} shows the distribution of $W_{50}$ versus virial velocity $V_{\rm vir}\propto m_{\rm vir}^{1/3}$, coloured by $c_{\rm vir}$, for a sample of $1000$ galaxies observed with a fixed inclination angle of $i=35^\circ$, with the \emph{left (right) panel} showing results for noiseless (noisy) profiles. We see that $W_{50}$ at fixed $i$ is almost completely determined by $V_{\rm vir}$ and $c_{\rm vir}$: there is a tight correlation  between $W_{50}$ and $V_{\rm vir}$, with the scatter around the mean relation at fixed $V_{\rm vir}$ itself being quite tightly correlated with $c_{\rm vir}$ (we measure Spearman correlation coefficients between $W_{50}$ and $c_{\rm vir}$ of $\gtrsim0.2$ in bins of $V_{\rm vir}\gtrsim100\,\kms$, rising to nearly $\sim0.8$ at $V_{\rm vir}\gtrsim300\,\kms$). The trends seen are also consistent with the $m_{\rm vir}$-$c_{\rm vir}$ degeneracy discussed earlier in the context of mass-modelling. A comparison between the two panels shows that the effect of the chosen level of noise is minimal (see also appendix~\ref{app:templatematching}).

Fig.~\ref{fig:mock-W50} is formatted identically to Fig.~\ref{fig:mock-W50-fixedincl} and shows results for the same 1000 galaxies now oriented randomly (i.e., $\sin i$ uniformly sampled as in the bottom panel of Fig.~\ref{fig:mock-lineprofiles}). Randomising the inclinations clearly has a substantial effect, with a large scatter between $W_{50}$ and $V_{\rm vir}$ and a correspondingly weaker correlation between $W_{50}$ and $c_{\rm vir}$ at fixed $V_{\rm vir}$ (Spearman correlation coefficients now drop to $\lesssim0.15$ over nearly the entire range of $V_{\rm vir}$). We return to a discussion of inclination effects in the context of the ability to constrain model variations in section~\ref{sec:results}.

\subsection{Realistic samples}
\label{subsec:W50-realistic}
\noindent
In order to be useful as a probe of small-scale ($\lesssim10\kpch$) physics, it is important that variations in the \Hi\ velocity width function be robust to observational systematics and errors. We therefore turn to constructing samples that mimic actual surveys such as ALFALFA \citep{giovanelli+05,giovanelli+07}. 

\begin{figure*}
\centering
\includegraphics[width=0.85\textwidth,trim=8 10 5 5,clip]{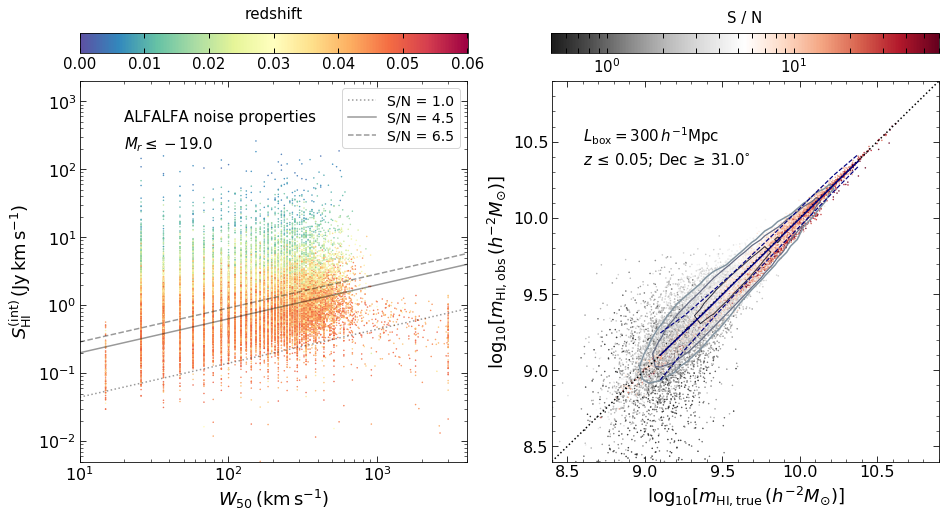}
\caption{{\bf Mock HI survey.} \emph{(Left panel:)} Mock observations of \Hi\ disks for a central galaxy sample selected by optical luminosity $M_r\leq-19$ and having $m_{\Hi}>0$, for the WS survey configuration chosen from a $(300\Mpch)^3$ simulation volume ($z\leq0.05$ and ${\rm Dec}\geq31^\circ$, survey area $\sim10,000\,{\rm deg}^2$) containing $\sim27,400$ galaxies. The disks are observed with randomised inclination angles assuming a $3.5^{\prime}$ beam width and velocity channel width and r.m.s. noise appropriate for the ALFALFA survey (see appendix~\ref{app:SN}). Each marker shows the integrated flux density $S_{\Hi}^{\rm (int)}$ and velocity width $W_{50}$ of an individual galaxy, computed using the algorithm outlined in section~\ref{subsec:W50-realistic} and coloured by the galaxy redshift \citep[c.f., e.g., fig.~1 of][]{martin+10}. Solid black line shows the threshold ${\rm S/N} = 4.5$ used in the subsequent analysis, calculated using \eqn{eq:S/N}. Dashed line shows the S/N threshold of 6.5 used in ALFALFA analyses, while the dotted line shows the value ${\rm S/N}=1$ for reference. \emph{(Right panel:)} Quality of mass recovery as a function of S/N. Markers show the value of $m_{\Hi}$ inferred from the observations in the left panel using \eqn{eq:mHI-SHI} against the true $m_{\Hi}$ value for each galaxy, coloured by the S/N of the observation. Gray lines show contours of equal number, with levels of $(100,200,800)$ objects on a grid of pixel width $0.083$ dex. Purple solid and dashed lines show the median and central $68\%$ region of $m_{\Hi,{\rm obs}}$ in bins of $m_{\Hi,{\rm true}}$. Dotted black line shows the 1:1 relation. }
    \label{fig:alfalfa-sampleprops}
\end{figure*}

In the context of our mock profiles, this requires (i) setting the velocity channel width $\Delta v$ and noise per channel $\sigma_{\Delta v}$ to  values matching the required survey, (ii) processing the resulting noisy profile of each mock galaxy using a realistic template fitting procedure and (iii) calculating a signal-to-noise ratio S/N. The sample can then be constructed using a threshold on S/N. We use the following method to create an observed catalog of $S_{\Hi}^{(\rm int)}$ and $W_{50}$ values, with details provided in appendix~\ref{app:SN}.
\begin{itemize}
\item For each mock galaxy in the chosen sample, we produce a noise-free \Hi\ velocity profile as detailed in section~\ref{subsec:vrot-Hivp}, with the channel width $\Delta v$ set by \eqn{eq:Dv-value} evaluated at $z=0$.
\item We add independent Gaussian noise to each channel with width $\sigma_{\Delta v}$ from \eqn{eq:sigDv-value}.
\item We smooth each profile using a 3-point Hann filter, which takes value $0.5$ at the central channel and value $0.25$ at each adjacent channel, being zero thereafter.
\item We  apply the template-matching procedure of appendix~\ref{app:templatematching} and estimate $W_{50}$ as the width at half the peak height of the (symmetric) best-fitting template for each noisy, smoothed profile.
\item Knowing $W_{50}$, a spectral extent $\Delta W$ is set using \eqn{eq:DW-def} and we estimate $S_{\Hi}^{(\rm int)}$ for each object by integrating  the smoothed profile over the range $v\in(-\Delta W/2,\Delta W/2)$ relative to the systemic velocity. We \emph{do not} introduce errors in determining the systemic velocity, instead using the true value as produced by our mock algorithm. 
\item The S/N  is then calculated using \eqn{eq:S/N}. 
\end{itemize}

Having generated a set of noisy measurements of $S_{\Hi}^{(\rm int)}$ and $W_{50}$ from each mock profile, we implement the 2-dimensional step-wise maximum likelihood (2DSWML) technique \citep[e.g.,][]{eep88} as described by \citet[][see their appendix~B]{martin+10} to infer the joint distribution $\phi_{\rm 2d}(\log[m_{\Hi}],\log[W_{50}])$ of \Hi\ mass and velocity width. Briefly, the maximum likelihood solution for the shape of the 2-dimensional density of galaxies $\phi_{mw}$ in bins of log-mass (labelled by $m$) and log-width (labelled by $w$) takes the form
\be
\phi_{mw} = N_{mw}/\sum_i\left(\frac{H_{imw}}{\sum_{m^\prime,w^\prime}H_{im^\prime w^\prime}\phi_{m^\prime w^\prime}}\right)\,,
\label{eq:2dswml-soln}
\ee
Here $N_{mw}$ is the observed galaxy count in the 2d bin, $\sum_i$ indicates a sum over all galaxies and $H_{imw}$ is the `completeness matrix' defined as
\be
H_{imw} = \frac1{\Delta m\Delta w}\int_{w^-}^{w^+}\der\tilde w \int_{m^-}^{m^+}\der\tilde m\,\Cal{C}_i(\tilde m,\tilde w)\,,
\label{eq:compmat}
\ee
where $(w^-,w^+)$ and $(m^-,m^+)$ indicate the bin edges, $\Delta w$ and $\Delta m$ are the corresponding bin widths and the completeness function $\Cal{C}_i(m,w)$ for the redshift of the $i^{\rm th}$ galaxy is unity if the $S/N$ returned by \eqn{eq:S/N} using this redshift and the mass-width pair $(m,w)$ exceeds the chosen threshold $(S/N)_{\rm min}$, and is zero otherwise. In practice, due to our standardised choice of spectral extent for defining $S/N$, $H_{imw}$ can be written in closed form as a function of $D_{\rm L}(z_i)$, $m$ and $w$ (parametrised by survey-dependent quantities such as channel width and noise r.m.s.). Equation~\eqref{eq:2dswml-soln} is then iterated to obtain a convergent solution for $\phi_{mw}$ (we have found that 10 iterations are more than sufficient). 

Since \eqn{eq:2dswml-soln} is insensitive to the normalisation of $\phi_{mw}$, this is fixed as follows \citep[appendix B1 of][]{martin+10}. We first normalise $\phi_{mw}$ to unity, such that $\Delta m\Delta w\sum_{m,w}\phi_{mw}=1$. We then estimate the number density of objects in the survey, accounting for survey incompleteness, using
\be
n_{\rm sur} = V_{\rm sur}^{-1} \sum_i\frac1{\sum_{m,w}H_{imw}\phi_{mw}}\,,
\label{eq:nsur}
\ee
where $V_{\rm sur}$ is the survey volume. Finally, the required 2d number density $\phi_{\rm 2d}(\log[m_{\Hi}],\log[W_{50}])$ is estimated as the product of $n_{\rm sur}$ and the unit-normalised $\phi_{mw}$.
Integrating $\phi_{\rm 2d}(\log[m_{\Hi}],\log[W_{50}])$ over $\log[W_{50}]$ gives the \Hi\ mass function, while integrating over $\log[m_{\Hi}]$ gives the \Hi\ velocity width function.

\section{Results}
\label{sec:results}
\noindent
In this section, we present the results of our algorithm for our default model as well as a number of variations. In the following, we will use an ALFALFA-like survey configuration selected from the ${\rm L}300\_{\rm N}1024$ box by placing the observer at the center of one box face (see appendix~\ref{app:rsd}) and selecting galaxies satisfying $z\leq0.05$ and ${\rm Dec}\geq31^\circ$ which gives a survey area of $\sim10,000\,{\rm deg}^2$ and a volume $\simeq(151\Mpch)^3$. (For comparison, the complete ALFALFA survey covers $\sim7000\,{\rm deg}^2$ with $z\lesssim0.05$.)  We select central galaxies having optical magnitude $M_r\leq-19$ (this is set by the resolution limit of the simulation box, see PCS21) and $m_{\Hi}>0$, which results in $\sim27,400$ galaxies. As before, we assume a telescope beam width of $\theta_{\rm beam} = 3.5^\prime$ matching the Arecibo value. Also, as in sections~\ref{subsec:vrot-Hivp} and~\ref{subsec:W50-basicprops}, we use $\sigma_v=10\,\kms$ for all galaxies in our default model. 

\subsection{Default model}
\label{subsec:results-default}
\noindent
The \emph{left panel} of Fig.~\ref{fig:alfalfa-sampleprops} shows the observed $S_{\Hi}^{\rm (int)}$ and  $W_{50}$ obtained using the procedure outlined in section~\ref{subsec:W50-realistic} on our default mock sample. Each marker shows the observation for an individual galaxy and is coloured by the galaxy's redshift. For reference, the black lines show various constant S/N values. We clearly see that low S/N objects preferentially occur at higher redshift, as expected, but otherwise span a wide range of velocity widths. The vertical streaks, particularly apparent at low $W_{50}$, reflect our choice of velocity channel width of $\Delta v\simeq5\kms$ (appendix~\ref{app:SN}).\footnote{The small clumping of low S/N galaxies near $W_{50}\simeq3000\kms$ is due to a numerical choice in our analysis in which we only simulate \Hi\ profiles over the range $\pm1500\kms$ on either side of the object's systemic velocity.}
Below, we use the threshold ${\rm S/N}\geq4.5$ when constructing samples for estimating the velocity width function \citep[for comparison, ALFALFA analyses such as that of][typically use a threshold of 6.5]{martin+10}. We have checked that our results for \Hi\ abundances below are insensitive to small variations in this choice.

The \emph{right panel} of Fig.~\ref{fig:alfalfa-sampleprops} compares the \Hi\ mass $m_{\Hi,{\rm obs}}$ estimated from the observed $S_{\Hi}^{\rm (int)}$ using \eqn{eq:mHI-SHI} (replacing the integral on the right hand side with $S_{\Hi}^{\rm (int)}$) with the true mass $m_{\Hi,{\rm true}}$ from the mock catalog. Each marker is coloured by the S/N. We see that large departures from the $1:1$ relation (dotted black line) occur predominantly at low S/N. This is further quantified by the blue solid and dashed lines, which respectively show the median and central $68\%$ region of $m_{\Hi,{\rm obs}}$ in bins of $m_{\Hi,{\rm true}}$: the solid line closely follows the $1:1$ relation while the dashed lines enclose a narrow region at high mass, which broadens towards lower masses where the fraction of low S/N observations is higher.

\begin{figure}
\centering
\includegraphics[width=0.48\textwidth,trim=8 10 10 7,clip]{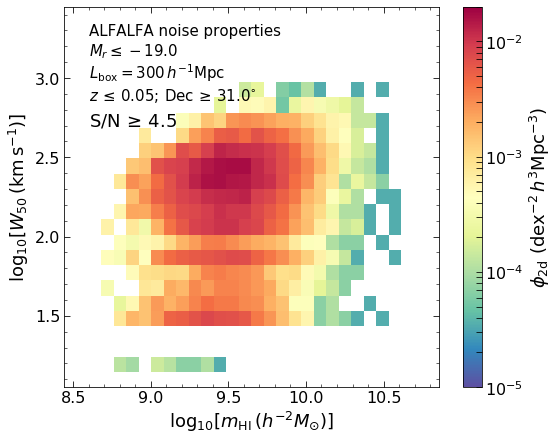}
\caption{{\bf Galaxy abundances.} 2-dimensional abundance $\phi_{\rm 2d}(\log[m_{\Hi}],\log[W_{50}])$ inferred from the data in the left panel of Fig.~\ref{fig:alfalfa-sampleprops} using the 2DSWML method.}
\label{fig:alfalfa-2dabundance}
\end{figure}

\begin{figure*}
\centering
\includegraphics[width=0.9\textwidth]{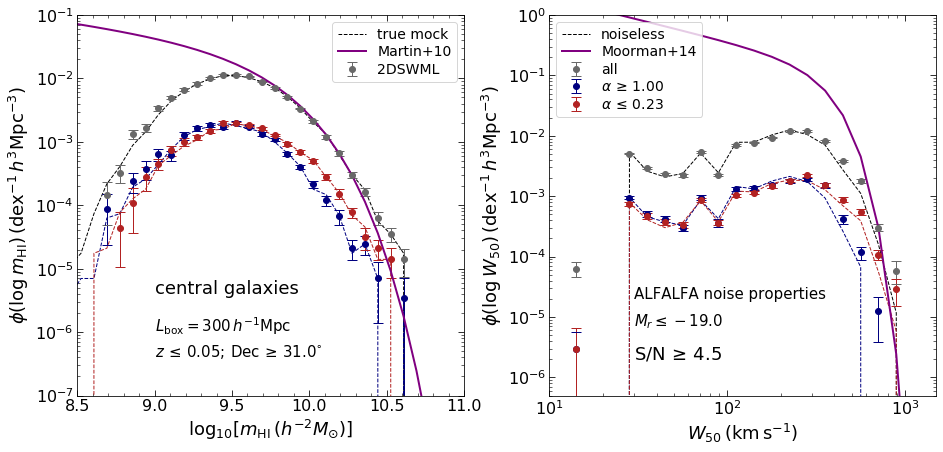}
\caption{{\bf Galaxy abundances.}
\Hi\ mass function \emph{(left panel)} and velocity width function \emph{(right panel)} of central galaxies, calculated as the integral of the 2-d abundance $\phi_{\rm 2d}(\log[m_{\Hi}],\log[W_{50}])$ from Fig.~\ref{fig:alfalfa-2dabundance} over $\log[W_{50}]$ and $\log[m_{\Hi}]$, respectively (gray symbols with error bars). 
Error bars were computed by applying the 2DSWML method separately to $50$ bootstrap samples and taking the standard deviation of the resulting abundances.
Dotted black lines show the underlying true distributions of $m_{\Hi}$ and $W_{50}$, computed by histogramming the true $m_{\Hi}$ values in the mock and estimates of $W_{50}$ from the noiseless velocity profiles.
Solid purple curves show the respective Schechter function fits from \citet{martin+10} and \citet{moorman+14} using the $\alpha0.4$ ALFALFA sample.
Blue (red) symbols with errors show the abundances for galaxies chosen to reside in  anisotropic (isotropic) tidal environments defined by the tidal anisotropy variable $\alpha$. Blue (red) dotted curves show the corresponding true distributions calculated similarly to the dotted black curves.}
\label{fig:alfalfa-abundances}
\end{figure*}

Fig.~\ref{fig:alfalfa-2dabundance} shows the 2-dimensional distribution $\phi_{\rm 2d}(\log[m_{\Hi}],\log[W_{50}])$ estimated from these observations using the 2DSWML method, i.e., after correcting for the incompleteness caused by the ${\rm S/N}\geq4.5$ threshold. There is a weak but distinct bimodality in the distribution along the $W_{50}$ direction, with a prominent excess around $W_{50}\simeq250\kms$ and a somewhat smaller excess near $W_{50}\simeq30\kms$, below which the distribution truncates sharply. This feature  could be partly due to the incompleteness inherent in our base sample caused by the optical selection of $M_r\leq-19$. This systematically misses \Hi-bearing galaxies progressively smaller than $m_{\Hi}\lesssim10^{9.7}\Mhsq$ (PCS21; see also below) and cannot be accounted for by the 2DSWML technique. So, e.g., it is possible that the missing galaxies would preferentially occupy widths $W_{50}\sim10^{1.5}$-$10^2\kms$, thus filling in the decrement between the two maxima. We see, however, that the bimodality in $W_{50}$ persists even when focusing  on galaxies with $m_{\Hi}\geq10^{9.7}\Mhsq$, and also in the absence of noise (not shown), indicating that this may be a genuine feature of the model.

This is explored further in Fig.~\ref{fig:alfalfa-abundances} which shows the integrals over this 2-d distribution to yield the \Hi\ mass function \emph{(left panel)} and velocity width function \emph{(right panel)} as the gray points with errors, compared with the respective noiseless distributions in the mock shown as the dotted black lines. This comparison shows that the 2DSWML method accurately recovers the underlying distribution of $m_{\Hi}$ and $W_{50}$, except perhaps at the largest $W_{50}$ where the abundance is overestimated compared to the noiseless case, and the smallest $W_{50}$ where some spurious counts are recorded. (The error bars were computed by applying the 2DSWML method to each of 50 bootstrap samples and taking the standard deviation of the resulting 1-d distributions.) To assess the level of incompleteness relative to actual ALFALFA observations, we show Schechter function fits to $\phi(m_{\Hi})$ and $\phi(W_{50})$ (solid purple curves) as calibrated by \citet{martin+10} and \citet{moorman+14}, respectively. For $\phi(m_{\Hi})$, we reproduce the result alluded to above \citep[see][for a detailed discussion]{pcp18} that the \Hi\ mass function produced by the PCS21 algorithm is incomplete for $m_{\Hi}\lesssim10^{9.7}\Mhsq\simeq1.12M_\ast$, where $M_\ast=10^{9.65}\Mhsq$ is the knee of the \citet{martin+10} Schechter fit to $\phi(m_{\Hi})$. The distribution of $W_{50}$, on the other hand, clearly suffers more than that of $m_{\Hi}$ from this inherent incompleteness of our mocks. We see that $\phi(W_{50})$ is only complete for $W_{50}\gtrsim700\kms\simeq1.84W_\ast$, with $W_\ast=380\kms$ being the knee of the \citet{moorman+14} Schechter fit to $\phi(W_{50})$. The bimodality in the $W_{50}$ distribution mentioned above is apparent, although somewhat suppressed, in the right panel of Fig.~\ref{fig:alfalfa-abundances} where $\phi(W_{50})$ traced out by the gray points shows a shallow minimum around $W_{50}\simeq50\kms$.

Of course, since our mocks are fundamentally limited by the resolution of the underlying HOD, an apples-to-apples comparison would require comparing them with optically selected subsamples of the ALFALFA survey. Alternatively, one could compare estimates of the conditional distribution $\phi(W_{50}|m_{\Hi}>10^{9.7}\Mhsq)$ for which our mocks are expected to produce complete results. 
Another option would be to explore AM techniques to access the low-$m_{\Hi}$ regime. We leave such comparisons for future work.

\subsection{Variations}
\label{subsec:results-vary}
\noindent
Our primary motivation in studying \Hi\ velocity profiles was to investigate their potential in constraining the baryon-dark matter connection in the $\Lambda$CDM framework. The results of section~\ref{subsec:vrot-Hivp} suggest that $\phi(W_{50})$ is likely to be sensitive to correlations involving inclination, disk size, halo mass, concentration and, to a lesser extent, the physics of quasi-adiabatic relaxation and the intrinsic width of the \Hi\ 21 cm line (see Figs.~\ref{fig:HIvp-NGC99-vary} and~\ref{fig:HIvp-UGC00094-vary}). In this section, we study the effect of such correlations on the shape of $\phi(W_{50})$.

\subsubsection{Sensitivity to environment}
\label{subsubsec:results-env}
\noindent
All galaxy properties (except $\sigma_v$) in our default model are ultimately related to the mass of the host halo through the underlying HOD. Since halo mass correlates with environment, it is worth asking what the model predicts for the environment dependence of the \Hi\ observables. 

The cosmic web environment of galaxies or their host haloes can be defined in a number of ways. While the large-scale overdensity of dark matter is perhaps the most commonly used discriminator of environment \citep[e.g.,][]{as07,goh+19}, recent work has emphasized the importance of the local tidal anisotropy in explaining many environmental trends of dark matter haloes \citep{hahn+09,bprg17,phs18a,rphs19}. The red (blue) markers in Fig.~\ref{fig:alfalfa-abundances} show abundances for galaxy samples selected by low (high) values of the halo-centric tidal anisotropy parameter $\alpha$, which is inherited by each galaxy from its host halo and is defined at a scale $\sim4\times$ the host radius $R_{\rm 200b}$. We refer the reader to \citet{phs18a} for a detailed definition of $\alpha$ (see their equation~10) and a description of how it is measured in an $N$-body simulation, but only note here that values $\alpha\gtrsim0.5$ correspond to haloes in filamentary environments while $\alpha\lesssim0.2$ corresponds to node-like environments (which could occur for massive objects at the intersection of large filaments or low-mass, isolated objects in voids). The base sample from which these subsamples are created is the same S/N thresholded set of galaxies used for producing the gray markers in Fig.~\ref{fig:alfalfa-abundances}.

Our chosen thresholds $\alpha\leq0.23$ and $\alpha\geq1$ lead to subsamples of approximately equal number ($\sim2500$) before applying the S/N threshold. We see that there is a distinct difference between the two subsamples at both, large $m_{\Hi}$ and large $W_{50}$, with the abundance of objects in filamentary environments being suppressed in each case. We can understand this as an effect of halo mass: filamentary haloes with high $\alpha$ tend to span a range of lower halo mass than node-like haloes which exist in all mass ranges, with massive haloes residing almost exclusively in low-$\alpha$ environments \citep[see, e.g., fig.~7 of][]{phs18a}. The suppression of abundances in filamentary environments is then a natural consequence of the correlation between velocity width and halo mass (see Fig.~\ref{fig:mock-W50}).
At low $m_{\Hi}$ and especially at low $W_{50}$, we see that the environmental cuts leave essentially no imprint on the abundances, apart from the obvious decrease due to reduced overall numbers.

We also repeated this exercise after splitting samples by the value of $\delta_{2\Mpch}$, the halo-centric dark matter density contrast, smoothed with a Gaussian filter of radius $2\Mpch$. Upon choosing high and low thresholds $\delta_{2\Mpch}$ that give subsamples of approximately the same size as the $\alpha$-split subsamples (i.e., $\sim2500$ objects before applying the S/N threshold), we found that the resulting abundances of galaxies with high (low) $\delta_{2\Mpch}$ are \emph{quantitatively} very similar to those of galaxies with low (high) $\alpha$. To avoid clutter, we have not separately shown these results in Fig.~\ref{fig:alfalfa-abundances}. This similarity can be understood from the fact that, (i) there is a strong positive correlation between $\alpha$ and $\delta_{2\Mpch}$ (Spearman correlation of $\simeq0.62$ for the sample shown by the gray markers in Fig.~\ref{fig:alfalfa-abundances}) and (ii) these environmental trends are ultimately derived from halo mass alone in our default model.

It will be very interesting to confront these predictions with corresponding observational results. Recently, \citet{moorman+14} have reported results for the \Hi\ mass function and velocity width function in ``void-like" and ``wall-like" environments. This environmental classification was based on the void catalog constructed by \citet{pan+12} which used the Void Finder algorithm of \citep{ep97,hv02} in which wall galaxies are first identified based on a nearest neighbour criterion and voids are then constructed by growing empty spheres in the wall-galaxy sample. At $W_{50}\gtrsim300\kms$, void-like environments show a suppression in the velocity width function relative to wall-like environments \citep[fig.~9 of][]{moorman+14}, qualitatively in agreement with the difference between the blue and red points in Fig.~\ref{fig:alfalfa-abundances} which correspond to low- and high-density environments, respectively. As mentioned above, an apples-to-apples comparison would require observational samples selected by optical properties, and also require using the same definitions of environment in both mocks and data, which we defer to future work.

\subsubsection{Sensitivity to relaxation physics}
\label{subsubsec:results-relax}
\noindent
As discussed in section~\ref{subsec:vrot}, the quasi-adiabatic relaxation physics of dark matter in each host halo is parametrised by the quantity $q_{\rm rdm}$, whose default value is set to $q_{\rm rdm}=0.68$. We also saw in section~\ref{subsubsec:varyparams} that changing $q_{\rm rdm}$ has relatively small effects as compared to other variables, but that these effects arise from a complex combination of dark matter and baryonic variables. In this section, we study the predicted effects of these changes on $\phi(W_{50})$.\footnote{The mass function $\phi(m_{\Hi})$ is, by construction, totally insensitive to $q_{\rm rdm}$ in our model.}

\begin{figure}
\centering
\includegraphics[width=0.46\textwidth]{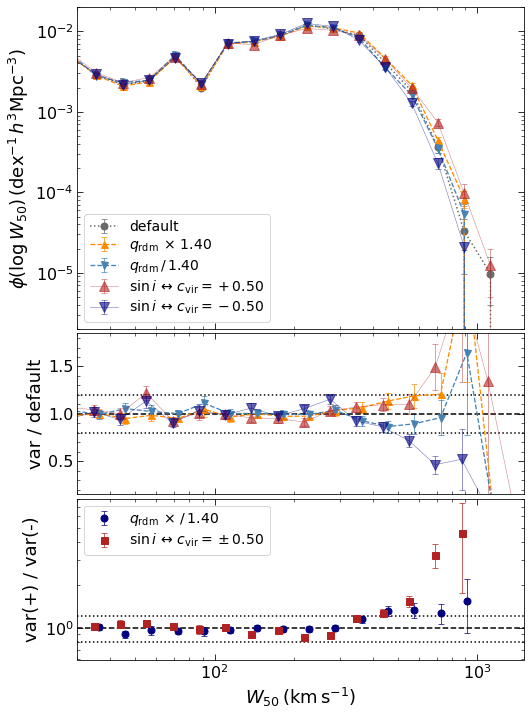}
\caption{{\bf HI velocity width function in alternative models.} \emph{(Top panel:)} $\phi(\log[W_{50}])$ for the default model (gray circles joined with dotted line, repeated from the right panel of Fig.~\ref{fig:alfalfa-abundances}) compared with results when varying the relaxation parameter $q_{\rm rdm}$ (small triangles joined by dashed lines, described in section~\ref{subsubsec:results-relax}), or including a correlation between inclination $i$ and halo concentration $c_{\rm vir}$ (large triangles joined by solid lines,  described in section~\ref{subsubsec:results-cvirincl}). For each alternative model, upward (downward) variations of the  relevant parameter  are shown  using upward (downward) pointing triangles with warmer (cooler) colours, and used galaxy samples defined identically to the one used  for the default model (see Fig.~\ref{fig:alfalfa-abundances}).
\emph{(Middle panel:)} Ratio of abundances in the alternative models to those in the default model, formatted identically to  the \emph{top panel}.
\emph{(Bottom panel:)} Ratio of abundances in the upward and downward parameter variations for each alternative model.  Circles and squares respectively show the results when varying $q_{\rm rdm}$ and the $\sin\,i\leftrightarrow c_{\rm vir}$ correlation.
Horizontal dotted lines in the \emph{middle} and \emph{bottom panels} indicate $\pm20\%$ deviations around unity (horizontal dashed line). Error bars in the \emph{top panel} were estimated  using $50$ bootstrap samples for each case, as in Fig.~\ref{fig:alfalfa-abundances}, while those in the \emph{middle} and \emph{bottom} panels were estimated using error propagation.
For $W_{50}\gtrsim300\kms$, we see $\sim20\%$ effects when varying $q_{\rm rdm}$ and up  to factor $\sim5$ effects when varying the $\sin\,i\leftrightarrow c_{\rm vir}$ correlation. For lower $W_{50}$, neither of the variations leads to any significant effect. 
The variations involving $\sigma_v$ (section~\ref{subsubsec:results-sigvSigHI}) and the correlation $h_{\Hi}\leftrightarrow c_{\rm vir}$ (section~\ref{subsubsec:results-cvirhdisk}) do not lead to any significant effect and are therefore not shown.
}
\label{fig:abundance-ratios}
\end{figure}

We have repeated the procedure outlined in section~\ref{subsec:W50-realistic} for two variations around the default model, setting $q_{\rm rdm} = 0.68\times1.4 \simeq 0.95$ in one and $q_{\rm rdm} = 0.68/1.4 \simeq 0.49$ in the other (the same as used in Figs.~\ref{fig:HIvp-NGC99-vary} and~\ref{fig:HIvp-UGC00094-vary}). The larger value thus represents near-perfect angular mometum conservation, while the lower value is observationally interesting for the radial acceleration relation in the high-acceleration regime \citep{ps21}. Fig.~\ref{fig:abundance-ratios} shows the results for the velocity width function for the ALFALFA-like sample. We see that these variations lead to essentially no effect for $W_{50}\lesssim300\kms$, while larger widths show small but significant departures from the default model, with the difference between the upward and downward variation in $q_{\rm rdm}$ exceeding $\sim20\%$ for $W_{50}\gtrsim500\kms$ \emph{(bottom panel)}. 

In the context of the discussion in section~\ref{subsubsec:varyparams}, these trends would be understandable if, at low $m_{\rm vir}$ (and hence low $W_{50}$), our mock galaxies had stellar masses $m_\ast$ that were preferentially above the AM relation used in Figs.~\ref{fig:HIvp-NGC99-vary} and~\ref{fig:HIvp-UGC00094-vary}, while at high $m_{\rm vir}$ (high $W_{50}$) the mock $m_\ast$ values were preferentially lower than the AM value. As we saw there, a low-$m_{\rm vir}$ halo with a larger-than-AM $m_\ast$ would be much less sensitive to $q_{\rm rdm}$ than a high-$m_{\rm vir}$ halo with a lower-than-AM $m_\ast$. Indeed, the stellar mass incompleteness induced by our intrinsic luminosity threshold of $M_r\leq-19$ leads to exactly such an effect: fig.~12 of PCS21 shows that galaxies with $m_{\rm vir}$ lower (higher) than $\sim10^{11.6}\Mh$ have $m_\ast$ values preferentially substantially above (slightly below) the AM relation. We conclude that the lack of sensitivity of the width function to $q_{\rm rdm}$ at low $W_{50}$ is likely due to the stellar mass incompleteness of our sample.

\subsubsection{Correlation between gas surface mass density and intrinsic width}
\label{subsubsec:results-sigvSigHI}
\noindent
We saw in Figs.~\ref{fig:HIvp-NGC99-vary} and~\ref{fig:HIvp-UGC00094-vary} (upper middle-right panels) that the shape of the \Hi\ velocity profile responds in a small but distinctive manner to the value of the intrinsic dispersion $\sigma_v$. Namely, increasing (decreasing) $\sigma_v$ makes the individual horns broader (sharper). Since our default model used the constant $\sigma_v=10\kms$, it is interesting to ask whether variations in $\sigma_v$ might leave an imprint in $\phi(W_{50})$ or related quantities. Observationally, while early work using small galaxy samples indicated that $\sigma_v$ is remarkably insensitive to galaxy properties \citep{sb99}, later work has revealed strong correlations between $\sigma_v$ and variables such as the surface density of \Hi\ mass ($\Sigma_{\Hi}$), of stellar mass ($\Sigma_{\rm cgal}$) or of baryonic mass ($\Sigma_{\rm bary}$) \citep[e.g.,][]{stilp+13}. Such correlations might be connected to the physics of supernova feedback, although this is not a settled question as yet \citep[see, e.g.,][]{ubf19,bacchini+20}.

With this motivation, we have therefore explored the following variations around our default model: (a) setting $\sigma_v=8\kms$ and (b) setting $\sigma_v$ as a Gaussian distributed  variable with mean $8\kms$ and standard deviation $2\kms$, perfectly correlated or anti-correlated with the surface density $\Sigma_{\Hi}$. In practice, for variation (b), we note that the \Hi\ disk scale $h_{\Hi}$ in the default model has a lognormal scatter of 0.06 dex around a median value $\avg{h_{\Hi}|m_{\Hi}}\propto m_{\Hi}^{0.5}$ at fixed $m_{\Hi}$ given by equation 8 of PCS21. Due to this, the surface density $\Sigma_{\Hi} \sim m_{\Hi}/h_{\Hi}^2$ in the default model has a lognormal scatter of 0.12 dex around a value independent of $m_{\Hi}$, with the scatter in $\log(\Sigma_{\Hi})$ being perfectly anti-correlated with that in $\log[h_{\Hi}/\avg{h_{\Hi}|m_{\Hi}}]$. To construct variation (b),  we therefore write $\sigma_v/(\kms) = 8 \mp 2\epsilon$, where $\epsilon = \log[h_{\Hi}/\avg{h_{\Hi}|m_{\Hi}}]/0.06$ is a standard normal deviate, with the minus (plus) sign leading to a perfect (anti-)correlation $\Sigma_{\Hi}\leftrightarrow\sigma_v$. The variation (a) tests the model's sensitivity to the absolute value of $\sigma_v$, while the variation (b) further tests for the  effect of a scatter in $\sigma_v$ as well as any strong (anti-)correlation with $\Sigma_{\Hi}$.

We found that $\phi(W_{50})$ for the ALFALFA-like sample  shows essentially \emph{no departure} (within errors) from the default model, for \emph{any} of  these variations. This is likely due to the fact that the changes we have explored in our $\sigma_v$ model are comparable to or smaller than the velocity sampling width (equation~\ref{eq:Dv-value}) of an ALFALFA-like survey. To avoid clutter, we have omitted these results from Fig.~\ref{fig:abundance-ratios}. Thus, while the shapes of individual \Hi\ profiles are affected by the value of $\sigma_v$, there is no observable imprint on $\phi(W_{50})$. We will see later, however, that beyond-width statistics describing the profile shape are, in principle, sensitive to these variations.

\subsubsection{Correlation between disk size and halo concentration}
\label{subsubsec:results-cvirhdisk}
\noindent
A potential correlation between disk size and halo concentration would be of great interest for galaxy formation models. As discussed by \citet{ps21}, a correlation between stellar bulge size and halo concentration, motivated by the size-spin correlations typically predicted by semi-analytical models \citep{mmw98,kravtsov13}, leads to interesting features in the radial acceleration relation. We have therefore investigated whether a similar correlation between $h_{\Hi}$ and $c_{\rm vir}$ leads to any effect in $\phi(W_{50})$. We follow \citet{ps21} and assume that the entire scatter of $0.06$ dex around the median \avg{h_{\Hi}|m_{\Hi}} in the distribution of $h_{\Hi}$ at fixed $m_{\Hi}$ is caused by variations in $c_{\rm vir}$, which allows us to write a modified model of disk sizes: $h_{\Hi} = \avg{h_{\Hi}|m_{\Hi}}\times(c_{\rm vir}/\avg{c_{\rm vir}|m_{\rm vir}})^{\pm0.375}$. Here \avg{c_{\rm vir}|m_{\rm vir}} is the median concentration at fixed halo mass, and the value of the exponent is fixed by noting that halo concentrations in our model obey a Lognormal distribution with a scatter of $0.16$ dex.

Interestingly, despite the strong effects of both $h_{\Hi}$ and $c_{\rm vir}$ on individual profiles (see Figs.~\ref{fig:HIvp-NGC99-vary} and~\ref{fig:HIvp-UGC00094-vary}), we found \emph{no significant effect} of this correlation on $\phi(W_{50})$, for either sign of the exponent, for the ALFALFA-like sample. We have checked that this absence of a signature in $\phi(W_{50})$ persists when binning galaxies by inclination (which could, in principle, be estimated from spatially resolved optical spectroscopy). To avoid clutter, we have not shown these results in Fig.~\ref{fig:abundance-ratios}. This lack of effect is likely due to the strong constraint of a small scatter in $h_{\Hi}$ at fixed $m_{\Hi}$, which our model treats as a purely observational input. An explanation of this small scatter in the  $\Lambda$CDM framework would therefore be an interesting avenue of future research.

\subsubsection{Correlation between inclination and halo concentration}
\label{subsubsec:results-cvirincl}
\noindent
The inclination angle of a galaxy relative to the observer is determined by the angular momentum vector of the rotating \Hi\ disk, which in turn is expected to correlate with the \emph{halo} angular momentum vector, which further correlates with local environment. Although each correlation in this chain is expected to be weak, this `intrinsic alignment' effect can, in principle, lead to an indirect correlation between inclination angles and halo properties such as concentration (since the latter also correlates with environment). We can ask whether the distribution of $W_{50}$ is sensitive to the amplitude of such a correlation.

We therefore introduce a correlation between $\sin\,i$ and $c_{\rm vir}$ (whose distribution is Lognormal, see above) by first drawing a Gaussian random variable $y = a\ln(c_{\rm vir}/\avg{c_{\rm vir}/m_{\rm vir}})+\epsilon$, where $\epsilon$ is a standard normal deviate uncorrelated with $c_{\rm vir}$. The values of $\sin i$ are then set by drawing uniform random numbers between zero and unity and rank ordering them according to the values of $y$. The constant $a$ is fixed so that the Spearman rank correlation coefficient $\sin i\leftrightarrow c_{\rm vir}$ takes some desired value: in  the following, we fix $\sin i\leftrightarrow c_{\rm vir} = \pm0.5$. Although this is large in magnitude compared to what one might expect in reality, it allows us to cleanly study the resulting trends in $\phi(W_{50})$.

We see in Fig.~\ref{fig:abundance-ratios} that this variation around the default model again leads to no effect in $\phi(W_{50})$ at $W_{50}\lesssim300\kms$, but shows large differences at higher widths. In particular, a positive (negative) correlation between inclination and halo concentration leads to larger (smaller) widths, with a corresponding increase (decrease) in the amplitude  of $\phi(W_{50})$. The results in Figs.~\ref{fig:HIvp-NGC99-vary} and~\ref{fig:HIvp-UGC00094-vary} show that these trends are sensible.

\section{Beyond-width statistics: excess kurtosis}
\label{sec:excesskurtosis}
\noindent
The qualitative similarity between the effects of a $\sin i\leftrightarrow c_{\rm vir}$ correlation and changes in the relaxation parameter $q_{\rm rdm}$ on the velocity width function make it interesting to study other aspects of the shape of \Hi\ velocity profiles. To this end, in this section we study the predicted distribution of the next most interesting shape statistic for symmetric profiles beyond the profile width, namely the \emph{excess kurtosis} $\kappa$.\footnote{For intrinsically asymmetric profiles, the \emph{skewness} derived from the third moment of the profile would also be interesting. Since the skewness vanishes for the symmetric profiles discussed in this work, we do not discuss it here.} We focus on noiseless profiles so as to  understand the intrinsic prediction of our default model and the variations discussed above, and comment later on the  requirements for measuring $\kappa$ observationally.

\begin{figure*}
\centering
\includegraphics[width=0.44\textwidth]{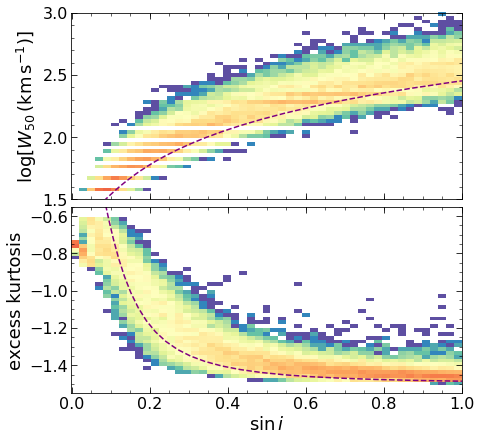}
\includegraphics[width=0.5\textwidth]{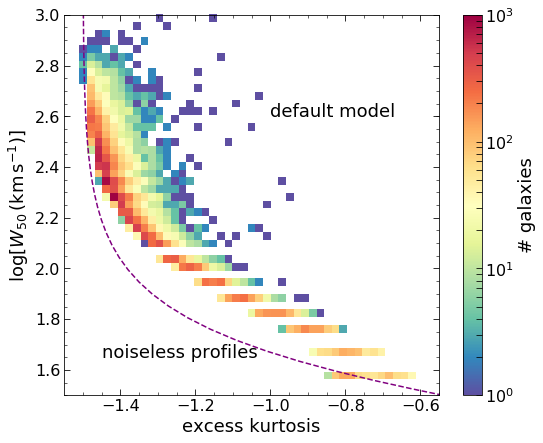}
\caption{{\bf Excess kurtosis, width and inclination}. Distributions of inclination $\sin i$ against velocity width $W_{50}$ \emph{(top left panel)} and excess kurtosis $\kappa$ \emph{(bottom left panel)} and joint distribution of $W_{50}$ and $\kappa$ \emph{(right panel)} for the \emph{noiseless profiles} generated using the default model for the same galaxies used in Fig.~\ref{fig:kurtosis}. 
The horizontal streaks along $W_{50}$ are due to the discrete channel width used in sampling each velocity profile (see also appendix~\ref{app:templatematching}).
The blue dashed curves show analytical approximations setting $\avg{v_{\rm rot}^2}^{1/2}=200\kms$ for the \emph{left panels} and $Y=1$ in \eqn{eq:kappa-analytical} for the \emph{right panel}. These qualitatively describe the main trends but are quantitatively different; the text discusses the implications of this comparison, especially for the tight relation seen in the \emph{right panel}.}
\label{fig:kurtosis-width-incl}
\end{figure*}

For a noiseless, symmetric velocity profile $S(v)$ which is centered at its systemic velocity, $\kappa$ can be written as
\be
\kappa\equiv c_4 / c_2^2 = \avg{v^4}/\avg{v^2}^2 - 3\,,
\label{eq:excesskurtosis-def}
\ee
where $\avg{v^n} \equiv \int\der v\,S(v)\,v^n/\int\der v\,S(v)$ is the $n^{\rm th}$ moment of the profile and $c_n$ is the $n^{\rm th}$ cumulant. A Gaussian-shaped profile would have $\kappa=0$ due to the vanishing of all $c_n$ with $n\geq3$. More generally, the assumption of symmetry and centering mean that $c_1 = 0 = c_3$, so that $c_2 = \avg{v^2}$ and $c_4 = \avg{v^4}-3c_2^2$, which  leads to the second equality. The expression in \eqn{eq:excesskurtosis-def} is equivalent to the usual definition of excess kurtosis as `kurtosis minus 3', where the kurtosis is defined as the ratio of the fourth central moment to the square of the variance. In general, a non-vanishing $\kappa$ is a measure of the relative importance of the tails of the profile as compared to a Gaussian shape \citep{westfall14}, with $\kappa < 0$ ($\kappa > 0$) indicating that the tails of the distribution are lighter (heavier) than that of a Gaussian.

\begin{figure}
\centering
\includegraphics[width=0.46\textwidth]{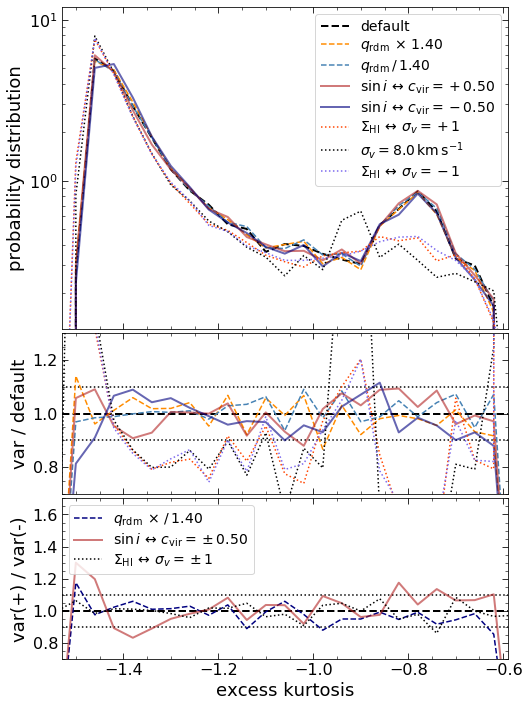}
\caption{{\bf Excess kurtosis distribution}. 
\emph{(Top panel:)} Probability distribution of excess kurtosis $\kappa$ (equation~\ref{eq:excesskurtosis-def}) for \emph{noiseless} \Hi\ velocity profiles of all objects in the ALFALFA-like mock sample shown in Fig.~\ref{fig:alfalfa-sampleprops}. 
Thick dashed black line shows the default model. 
Solid and thin dashed lines show the variations involving $q_{\rm rdm}$ and a $\sin i\leftrightarrow c_{\rm vir}$ correlation discussed in sections~\ref{subsubsec:results-relax} and~\ref{subsubsec:results-cvirincl}, respectively (see also Fig.~\ref{fig:abundance-ratios}). 
Dotted lines show the variations involving $\sigma_v$ discussed in section~\ref{subsubsec:results-sigvSigHI}. These include setting $\sigma_v=8\kms$ (dotted black) or distributed as a Gaussian with mean $8\kms$ and standard deviation $2\kms$ (red and blue dotted lines), with the red (blue) lines showing the case for perfect (anti-)correlation between $\sigma_v$ and the surface density of \Hi\ mass $\Sigma_{\Hi}$.
\emph{(Middle panel:)} Ratio of the $\kappa$ distribution for each variation with that in the default model.
\emph{(Bottom panel:)} For each variation involving $q_{\rm rdm}$, $\sin i\leftrightarrow c_{\rm vir}$ and $\Sigma_{\Hi}\leftrightarrow\sigma_v$, the curves show the ratio of the distribution in the upward variation to that in the downward variation.
Horizontal dotted lines in the \emph{middle} and \emph{bottom panels} indicate $\pm10\%$ deviations around unity (horizontal dashed line).
We see that only the variations involving $\sigma_v$ leave any noticeable imprint on the $\kappa$ distribution, which we discuss further in the main text.
Similarly to Fig.~\ref{fig:abundance-ratios}, the variation involving the correlation $h_{\Hi}\leftrightarrow c_{\rm vir}$ discussed in section~\ref{subsubsec:results-cvirhdisk} does not lead to any significant effect and is therefore not shown.
}
\label{fig:kurtosis}
\end{figure}

From \eqn{eq:SHI}, it is easy to show that the variance of $S_{\Hi}(v)$ can be written as 
$\avg{v^2} = \sigma_v^2(1+y^2/2)\approx (W_{50}/2)^2$, while $\kappa$ takes the form 
\be
\kappa = -\frac38\frac{y^4(2-Y)}{(1+y^2/2)^2}\,, 
\label{eq:kappa-analytical}
\ee
where we defined $y$ and $Y$ as
\be
y^2\equiv\avg{v_{\rm rot}^2}(\sin^2i)/\sigma_v^2\,;\quad Y\equiv\avg{v_{\rm rot}^4}/\avg{v_{\rm rot}^2}^2\,, 
\label{eq:yY-def}
\ee
with the averages appearing in $y$ and $Y$ being performed over the \Hi\ surface density, so that, e.g., $\avg{v_{\rm rot}^n} = \int_0^{r_{\rm max}}\der r_\perp\,r_\perp\,\Sigma_{\Hi}(r_\perp)\,v_{\rm rot}^n(r_\perp)/\int_0^{r_{\rm max}}\der r_\perp\,r_\perp\,\Sigma_{\Hi}(r_\perp)$.

We see that $\kappa<0$ always, provided $Y<2$. If the rotation curve $v_{\rm rot}(r_\perp)$ is in its flat part in the region where $r_\perp\,\Sigma_{\Hi}(r_\perp)$ has its support (i.e., near $r_\perp\simeq h_{\Hi}$), then $Y\simeq1$ and $\kappa$ becomes a function of $(W_{50}/\sigma_v)$ alone. In general, since we expect $\avg{v_{\rm rot}^2}\gg\sigma_v^2$, we will have $y\gg1$ except for nearly face-on galaxies. In this limit, which is where we expect most galaxies to be, $\kappa\to-(3/2)(2-Y)\simeq-3/2$, independent of inclination and nearly independent of $W_{50}$. For low-inclination galaxies such that $y\lesssim1$, $\kappa\propto-y^4(2-Y)$, thus becoming a strong function of both inclination and the intrinsic width $\avg{v_{\rm rot}^2}$.

Fig.~\ref{fig:kurtosis-width-incl} shows the joint distributions of $\kappa$, $W_{50}$ and $\sin i$ for the noiseless profiles in our default model,  using the ALFALFA-like mock sample shown in Fig.~\ref{fig:alfalfa-sampleprops}.\footnote{We remind the reader that our default model uses $\sigma_v=10\kms$ for all galaxies. Also, as in appendix~\ref{app:templatematching}, $W_{50}$ for each noiseless profile is directly estimated as the width at half its common peak height, without matching to any template.} We see all the trends discussed above. There is a tight and non-linear anti-correlation between $\kappa$ and $\log[W_{50}]$ \emph{(right panel)}, such that most galaxies are found near $\kappa\simeq-1.45$, with a smaller cluster near $\kappa\simeq-0.8$. The dashed purple line shows the approximation $Y\simeq1$ discussed above. While this broadly traces the $\kappa$-$W_{50}$ anti-correlation, it misses most of the distribution and has a different shape. This difference, as well as the scatter in the measured relation, can be attributed entirely to the fact that $Y\neq1$ for all galaxies in \eqn{eq:kappa-analytical}. The shape and scatter of the measured $\kappa$-$W_{50}$ relation, therefore, are potentially sensitive to the physics governing the distribution of $Y$.

The \emph{left panels} of Fig.~\ref{fig:kurtosis-width-incl} show that, as expected, both $\kappa$ and $W_{50}$ correlate with inclination at low values of $\sin i$, with $\kappa$ becoming nearly independent of inclination for $\sin i\gtrsim0.5$. The dashed purple curves in each panel show the prediction if we set $\avg{v_{\rm rot}^2}=(200\kms)^2$; this clearly provides a reasonable description of the qualitative trends. Since our mocks are incomplete at low $W_{50}$ (see Fig.~\ref{fig:alfalfa-abundances}), the structure and position of the $\kappa\simeq-0.8$ cluster of galaxies (which also all occur at the lowest $W_{50}$) is quite possibly not representative of an \Hi\ mass-complete sample, but should rather only be interpreted for an optical luminosity-complete sample with $M_r\leq-19$.

We now ask how sensitive the 1-dimensional $\kappa$ distribution is to variations around our default model, leaving a detailed study of the $\kappa$-$W_{50}$ relation to future work. The \emph{top panel} of Fig.~\ref{fig:kurtosis} shows the noiseless distribution $p(\kappa)$ for our default model (thick dashed black curve; same as integrating over $W_{50}$ in the right panel of Fig.~\ref{fig:kurtosis-width-incl}) and variations  (coloured lines). The bimodality mentioned above is now readily apparent. The variations around the default model we have explored mostly do not appear to affect $p(\kappa)$ substantially, as seen in the \emph{middle} and \emph{bottom panels} of Fig.~\ref{fig:kurtosis}, except for a clear dependence on the value of $\sigma_v$. In particular, the variation (a) from section~\ref{subsubsec:results-sigvSigHI} ($\sigma_v=8\kms$ instead of the default $10\kms$) leads to a shift in $p(\kappa)$ to lower values, with both modes being somewhat narrower than in the default case. The variation (b) from section~\ref{subsubsec:results-sigvSigHI} ($\sigma_v=8\pm2\kms$ with a Gaussian distribution) leads to even more interesting results. In this case, $p(\kappa)$ is identical to that of variation (a) for $\kappa\lesssim-1$, but has a distinctly broader high-$\kappa$ mode (compare the black dotted line with the red and blue dotted lines for $\kappa\gtrsim-0.9$). Finally, the \emph{sign} of the correlation between $\sigma_v$ and $\Sigma_{\Hi}$ does not lead to any noticeable difference (red and blue dotted lines are nearly identical, see also the \emph{bottom panel}). These effects of changing $\sigma_v$ are all naturally explained by \eqn{eq:kappa-analytical}, keeping in mind that decreasing $\sigma_v$ will increase $y$.

Finally, as regards observational estimates of $\kappa$, the presence of noise in realistic \Hi\ velocity profiles means that the integrals involved in measuring $\kappa$ in real data must be performed carefully. One approach would be to directly integrate the best-fitting templates obtained using the method outlined in appendix~\ref{app:SN}, provided the template shapes are flexible enough to capture the range of $\kappa$ seen in the noiseless profiles. The examples shown in Fig.~\ref{figapp:tempmatch} indicate that this would require the inclusion of at least $\Psi_4$, in addition to $\Psi_0$ and $\Psi_2$, in the Hermite function basis set used for building templates. Consequently, the least squares exercise would involve at least one more free parameter. We will explore the feasibility of this exercise, including the minimal requirements on the template basis functions, in future work.

\section{Conclusions}
\label{sec:conclude}
\noindent
We have studied the distribution of \Hi\ velocity profiles as measured by an observer in a $\Lambda$CDM universe, which constitutes a hitherto unexplored statistical probe of the small-scale baryon-dark matter connection.

As is well known, the velocity profile of an \Hi\ disk as seen by a distant observer can be derived using the galaxy's rotation curve (modulated by its observed inclination angle) and the mass distribution of \Hi\ in the disk \citep[e.g.,][see section~\ref{subsec:vrot-Hivp}]{sbr94}. Our analysis applied this calculation to the rotation curves of \Hi-bearing central galaxies having optical magnitude $M_r\leq-19$ in a statistically realistic mock catalog  of galaxies in a $(300\Mpch)^3$ box \citep[][hereafter, PCS21]{pcs21} constructed using an optical+\Hi\ halo occupation distribution (HOD) model \citep[][see section~\ref{sec:mock}]{pcp18,ppp19}. The HOD is constrained to reproduce the abundances and luminosity- and colour-dependent clustering of optically selected galaxies in SDSS, as well as the abundances and \Hi-dependent clustering of massive \Hi-selected galaxies in the ALFALFA survey. The rotation curves derived from the baryonified host haloes of these central galaxies have been shown to be in very good agreement with the median and scatter of the observed radial acceleration relation in the local Universe \citep{ps21}. 

We showed in section~\ref{subsec:examples} that, when constrained by observed \Hi\ profiles of nearby galaxies, along with knowledge of the disk inclination, our baryonification model produces realistic descriptions of their dark matter and baryonic content. Additionally, our model accounts for the quasi-adiabatic relaxation of dark matter in the presence of baryons in each halo. This suggests that our technique for generating \Hi\ disks could be a useful mass-modelling tool, particularly for objects with spatially resolved optical and radio spectra available. Our novel sample of \Hi\ velocity profiles, on the other hand (e.g., Fig.~\ref{fig:mock-lineprofiles}), and the resulting statistics  derived from our mock catalog by `observing' galaxies in redshift space (appendix~\ref{app:rsd})  constitute the first theoretical study of the statistical properties of velocity profiles in a $\Lambda$CDM universe. 

In addition to our default model for generating rotation curves and velocity profiles, we have explored a number of variations which could, in principle, affect the shapes of \Hi\ velocity profiles. These include changing the quasi-adiabatic relaxation physics (section~\ref{subsubsec:results-relax}), a correlation between gas surface mass density $\Sigma_{\Hi}$ and the \Hi\ intrinsic velocity dispersion $\sigma_v$ (section~\ref{subsubsec:results-sigvSigHI}), a correlation between \Hi\ disk size and halo concentration (section~\ref{subsubsec:results-cvirhdisk}), and a correlation between galaxy inclination and halo concentration (section~\ref{subsubsec:results-cvirincl}). 

A commonly used statistic derived from an \Hi\ velocity profile is its width $W_{50}$, which is sensitive to not only the galaxy's inclination but also other physical properties such as host halo mass and concentration (section~\ref{subsec:W50-basicprops}, Figs.~\ref{fig:mock-W50-fixedincl} and~\ref{fig:mock-W50}), as well as baryonic properties such as the \Hi\ disk size and intrinsic velocity dispersion (Figs.~\ref{fig:HIvp-NGC99-vary} and~\ref{fig:HIvp-UGC00094-vary}). Along with the \Hi\ mass function $\phi(m_{\Hi})$ \citep{zmsw05,martin+10}, the \Hi\ velocity width function $\phi(W_{50})$ is a natural product of large-volume surveys of \Hi-selected galaxies \citep{papastergis+11,moorman+14}, although it is only $\phi(m_{\Hi})$ which has been typically used for constraining models of galaxy evolution. In order to assess the constraining power of $\phi(W_{50})$, we therefore set up a realistic procedure for estimating $W_{50}$ by template-matching noisy \Hi\ velocity profiles measured in an ALFALFA-like survey (appendix~\ref{app:templatematching}) and consequently estimating $\phi(m_{\Hi})$ and $\phi(W_{50})$ using the 2DSWML method  (section~\ref{subsec:W50-realistic} and appendix~\ref{app:SN}). 
Our main results in this regard are as follows.

\begin{itemize}
\item For our default model, the 2DSWML method applied to ALFALFA-like noisy data accurately recovers the intrinsic $\phi(m_{\Hi})$ and $\phi(W_{50})$, except at $W_{50}\gtrsim400\kms$ where it overestimates $\phi(W_{50})$ and in the lowest $W_{50}$ bin where it returns a spurious count (Fig.~\ref{fig:alfalfa-abundances}; see also Fig.~\ref{fig:alfalfa-2dabundance}). 
\item Our default model for rotation curves, applied to a luminosity-complete mock catalog of central galaxies with $M_r\leq-19$, leads to an \Hi\ mass function that is complete for $m_{\Hi}\gtrsim10^{9.7}\Mhsq$ (\citealp{pcp18}; PCS21) but a velocity function that is complete only for $W_{50}\gtrsim700\kms\simeq1.84W_\ast$, where $W_\ast$ is the knee of the observed ALFALFA velocity width function (Fig.~\ref{fig:alfalfa-abundances}). As such, all our results should be interpreted for samples that are complete in optical luminosity rather than \Hi\ mass.
\item The default model (which is based on a `halo mass only' HOD) predicts distinct differences in $\phi(W_{50})$ for galaxies in tidally anisotropic (underdense) and isotropic (overdense) environments (blue and red symbols, respectively, in Fig.~\ref{fig:alfalfa-abundances}; see section~\ref{subsubsec:results-env} for a discussion).
\item Among the variations around the default model mentioned above, the strongest imprints on $\phi(W_{50})$ are seen when introducing a correlation between galaxy inclination and halo concentration, followed by variations in the quasi-adiabatic relaxation physics (Fig.~\ref{fig:abundance-ratios}). The effects of these variations are, however, degenerate with each other. The remaining variations showed no discernable effects on the $\phi(W_{50})$ for an ALFALFA-like survey.
\end{itemize}

We have also performed a preliminary study of beyond-width statistics, focusing on the excess kurtosis $\kappa$ (equation~\ref{eq:excesskurtosis-def}) of noiseless profiles of a luminosity-complete sample in an ALFALFA-like survey geometry, which led to the following conclusions.
\begin{itemize}
\item 
The analytical understanding of $\kappa$ (equation~\ref{eq:kappa-analytical}) predicts that $\kappa$ is negative and restricted to values $\gtrsim-1.5$, being a strong function of $W_{50}$ at any inclination, and of $\sin i$ at low inclination. This is borne out by Fig.~\ref{fig:kurtosis}.
\item The shape and scatter of the $\kappa$-$W_{50}$ relation are predicted to be sensitive to the distribution of the ratio $Y = \avg{v_{\rm rot}^4}/\avg{v_{\rm rot}^2}^2$ of \Hi-mass-weighted averages of galaxy rotation curves (section~\ref{sec:excesskurtosis}).
\item Among the variations around the default model, it is now the one involving changes in the intrinsic width $\sigma_v$ which leads to strong effects in the 1-dimensional $\kappa$ distribution at low inclinations, while the other variations lead to essentially \emph{no} effect (Fig.~\ref{fig:kurtosis}). The response of the $\kappa$-$W_{50}$ relation to such variations deserves further attention. The distribution of $\kappa$ could thus be a sensitive probe of baryonic physics in the turbulent \Hi\ disk, provided $\kappa$ can be robustly estimated from noisy profiles. Independent estimates of the inclination would make such analyses even more sensitive.
\end{itemize}

We end with a discussion of possible improvements and extensions of our model. Our analysis above was restricted to central galaxies, because it relies on the baryonification scheme described in  section~\ref{subsec:vrot} which has not yet been developed for the (subhalo) hosts of satellite galaxies. Indeed, our mocks do not use subhalo information from the $N$-body simulation at all, relying instead on empirical models for the spatial distribution and properties of (point-like) satellites (PCS21). Observationally, the clustering of \Hi-selected galaxies with projected separations $\lesssim300\kpch$ does require the inclusion of a small but significant number of \Hi-bearing satellite galaxies in groups \citep{guo+17,pcp18}. Such satellites are also likely to contain spatially disturbed distributions of \Hi\ due to tidal interactions with their dense environments and with other galaxies, possibly leading to preferentially asymmetric \Hi\ velocity profiles \citep{watts+20}. Tidal interactions would also strip away dark matter from a satellite's subhalo host, while interactions with the hot halo gas in massive groups can affect the star formation properties and gas content of the satellite  itself \citep[e.g.,][]{vdb+08}. All of these would affect the mass profile and hence rotation curve of the satellite, thus making it imperative to robustly model such effects using, e.g., subhalo demographics from high-resolution $N$-body experiments \citep[e.g.,][]{vdbtg05,jvdb16} along with empirical models for the stellar and \Hi\ spatial distribution. The modelling of satellite rotation curves, allowing for asymmetries such as warps in the \Hi\ distribution, is therefore a clear direction for future improvements in our model. The modelling of asymmetries in \Hi\ velocity profiles is, in general, an interesting avenue of research, although the statistical characterisation of asymmetry in observed samples, along with its connection to  galaxy properties, is yet to be settled  \citep[see, e.g.,][]{bbge19,wccp20,deg+20,watts+21}.

Our analysis above also did not fully exploit the spatial distribution of the galaxies in the surrounding cosmic web. It will be interesting to study the predictions of our model for clustering statistics such as mark correlations \citep{sheth05,skibba+13} using $W_{50}$ and/or $\kappa$ as marks. The presence of high-velocity clouds (HVCs) of \Hi\ due to substructure in the vicinity of an \Hi\ disk, which is currently not included in our model, could alter the shapes of individual \Hi\ velocity profiles, particularly in the tails \citep[e.g.,][]{sbr94}, and possibly also leave an imprint in clustering statistics. More generally, it would be interesting to develop compact summary statistics (e.g., using wavelet transformations) that can capture aspects of an individual \Hi\ velocity profile such as the shape of the individual horns, the height between each horn summit and the central trough, etc., which might be sensitive to the underlying baryonic and dark matter variables in different ways and therefore useful in breaking degeneracies.

Finally, weak gravitational lensing leaves a number of interesting signatures on the observed properties of rotating disks. In a spatially resolved galaxy spectrum, the axes along which the radial velocity is zero and maximum are perpendicular to one another if the object is not lensed.  The amount by which this angle differs from $90^\circ$ is a measure of the lensing signal \citep{blain02, morales06}.  Lensing will also modify the axis lengths of the image (while preserving surface brightness), producing an offset from the Tully-Fisher relation -- an effect known as Kinematic Lensing \citep{kl13}.  These are subtle effects that can be detected with even higher signal-to-noise if other photometric parameters (e.g., colour) are known \citep{cfss17}.  Our mock catalogs contain all the required spectroscopic and photometric information that is required to make realistic estimates of the strength of the expected signal from massive galaxies, simplifying the process of forecasting the constraints that HI surveys may place on the lensing potential \citep{ws21}. We will return to these ideas in future work.

\section*{Acknowledgments}
AP thanks Nishikanta Khandai for valuable discussions.
The research of AP is supported by the Associateship Scheme of ICTP, Trieste and the Ramanujan Fellowship awarded by the Department of Science and Technology, Government of India. 
TRC acknowledges support of the Department of Atomic Energy, Government of India, under project no.~12-R\&D-TFR-5.02-0700 and the Associateship Scheme of ICTP. 
This work made extensive use of the open source computing packages NumPy \citep{vanderwalt-numpy},\footnote{\url{http://www.numpy.org}} SciPy \citep{scipy},\footnote{\url{http://www.scipy.org}} Matplotlib \citep{hunter07_matplotlib},\footnote{\url{https://matplotlib.org/}} and Jupyter Notebook.\footnote{\url{https://jupyter.org}} 

\section*{Data Availability}
The mock catalogs generated by our algorithm will be shared upon reasonable request to the authors.

\bibliography{references}

\appendix
\section{Redshift space}
\label{app:rsd}
\noindent
Here we collect some relations that are useful when moving objects into redshift space and for determining observed redshifts based on local positions and velocities. Throughout, we consider a flat FLRW cosmology and assume that peculiar velocities are locally non-relativistic. Below, $z$ will generically denote redshift, $(X,Y,Z)$ will denote comoving Cartesian coordinates centered at the observer and $(v_X,v_Y,v_Z)$ will denote physical peculiar velocities relative to the Cartesian grid.

Consider a source at comoving distance $R$ from the observer, emitting at cosmic time $t$ corresponding to redshift $z=1/a(t)-1$ and observed at current epoch $t_0$. Let the source have a peculiar velocity $v_\parallel$ along the observer's line of sight. Then the light propagation integrals for two pulses separated by one wavelength $\lambda$ at the source are
\begin{align}
{\rm pulse\,1:}\quad& \int_0^{R}\der r = \int_t^{t_0}\frac{c\,\der t}{a(t)}\notag\\
{\rm pulse\,2:}\quad& \int_0^{R+v_\parallel\delta t}\der r = \int_{t+\delta t}^{t_0+\delta t_0}\frac{c\,\der t}{a(t)}\,
\label{appeq:lightprop}
\end{align}
where $\delta t = \lambda/c$ and $\delta t_0 = \lambda_{\rm obs}/c$, with $\lambda_{\rm obs}$ being the observed wavelength. Straightforward manipulation leads to the `cosmic Doppler' formula
\be
1+z_{\rm obs} \equiv \frac{\lambda_{\rm obs}}{\lambda} = \left(1+z\right)\left(1+\frac{v_\parallel}{c}\right)
\label{appeq:cosmicdoppler}
\ee
Consider now a cubic, periodic simulation box of comoving length $L_{\rm com}$ at cosmic time $t_{\rm sim}$ or redshift $z_{\rm sim}=1/a(t_{\rm sim})-1$. We wish to assign an `observed' redshift to a tracer (halo, galaxy, etc.) at a comoving position $\mathbf{r}_{\rm com}=(X,Y,Z)$ with peculiar velocity $\mathbf{v} = (v_X,v_Y,v_Z)$. Let us first do this using the so-called distant observer approximation and later generalise to arbitrary lines of sight.

\subsection{Distant observer approximation}
\label{app:RSpos:distob}
\noindent
Assume that the simulation box is sufficiently far from the observer along the Cartesian $Z$-direction, such that the comoving position vector $\mathbf{r}_{\rm com}$ of any tracer in the box relative to the observer satisfies $\mathbf{r}_{\rm com} = \hat n\,r_{\rm com}\approx \hat Z\,Z$. In other words, the line of sight $\hat n$ to \emph{any} tracer is approximately $\hat n = \hat Z$.

Let us write $Z = \bar Z + \delta Z$, where 
\be
\bar Z \equiv L_{\rm com}/2+ r_{\rm com}(z_{\rm sim}) \equiv L_{\rm com}/2+ \int_0^{z_{\rm sim}}\frac{c\,\der z}{H(z)}
\label{appeq:Zbar}
\ee
is essentially the comoving distance to redshift $z_{\rm sim}$ in the FLRW geometry, and $\delta Z$ is the actual comoving position of the tracer along the $Z$-direction in the simulation box, relative to the box center. We have chosen a convention in which the observer sits on one face of the box if $z_{\rm sim}=0$. We can then convert $\delta Z$ into a residual cosmic redshift $\delta z$ (in the absence of peculiar motion) using
\begin{align}
\delta Z 
&= \int_{z_{\rm sim}}^{z_{\rm sim}+\delta z}\frac{c\,\der z}{H(z)} -L_{\rm com}/2\notag\\
\Longrightarrow \delta z &\approx \frac{(1+z_{\rm sim})\left(\delta Z+L_{\rm com}/2\right)}{\ell_{\rm H,com}(z_{\rm sim})}\,,
\label{appeq:deltaZ-distob}
\end{align}
where the second line assumes that the box size $L_{\rm com}$ is much smaller than the comoving Hubble length 
\be
\ell_{\rm H,com}(z_{\rm sim}) = (1+z_{\rm sim})\,\ell_{\rm H}(z_{\rm sim}) \equiv \frac{c(1+z_{\rm sim})}{H(z_{\rm sim})}\,. 
\label{appeq:ellHcom}
\ee
Using this in the cosmic Doppler formula \eqref{appeq:cosmicdoppler} gives us the observed redshift of a tracer under the distant observer approximation (with the line of sight along the $Z$-direction)
\begin{align}
\frac{1+z_{\rm ob}}{1+z_{\rm sim}} &= \left(1 + \frac{(\delta Z+L_{\rm com}/2)}{\ell_{\rm H,com}(z_{\rm sim})}\right)\left(1+\frac{v_Z}{c}\right) \notag\\
&\approx 1 + \frac{(\delta Z+L_{\rm com}/2)}{\ell_{\rm H,com}(z_{\rm sim})} + \frac{v_Z}{c}\,.
\label{appeq:zob:distob}
\end{align}
For a simulation snapshot at $z_{\rm sim}=0$, this reduces to the familiar formula for the comoving redshift space position $\delta Z_{\rm S}$ along the line of sight: $\delta Z_{\rm S} = c\delta z/H_0-L_{\rm com}/2 = \delta Z + v_Z/H_0$.

\subsection{Arbitrary line of sight}
\label{app:RSpos:arbitlos}
\noindent
For a simulation box whose center is at $(0,0,\bar Z)$ relative to the observer, with $\bar Z$ (equation~\ref{appeq:Zbar}) not necessarily large, it is straightforward to show that the residual cosmic redshift $\delta z$ for a tracer at location $(\delta X,\delta Y,\delta Z)$ relative to the box center can be obtained by solving
\begin{align}
\delta R_{\rm S} &\equiv\left(\bar Z^2 + 2\bar Z\,\delta Z + \delta R^2\right)^{1/2} - \bar Z +L_{\rm com}/2 \notag\\
&= \int_{z_{\rm sim}}^{z_{\rm sim}+\delta z}\frac{c\,\der z}{H(z)}  \,,
\label{appeq:deltaRS}
\end{align}
where $\delta R^2\equiv\delta X^2 + \delta Y^2 +\delta Z^2$ and the first line defines the redshift space comoving distance residual $\delta R_{\rm S}$. The cosmic Doppler formula then becomes
\be
1+z_{\rm obs} = \bigg(1+z_{\rm sim} + \delta z(\delta R_{\rm S})\bigg)\left(1+\frac{\mathbf{v}\cdot\hat n}{c}\right)\,,
\label{appeq:cosmicdoppler-generic}
\ee
which assumes non-relativistic peculiar velocities but \emph{does not} assume a small box. Here $\delta z(\delta R_{\rm S})$ must be obtained by inverting \eqn{appeq:deltaRS} and the line of sight direction $\hat n$ is given by 
\be
\hat n = (\delta X,\delta Y,\bar Z+\delta Z)/(\bar Z-L_{\rm com}/2+\delta R_{\rm S})\,.
\label{appeq:los}
\ee
As a limiting case, we can recover the distant observer approximation by setting $\delta X=0=\delta Y$, so that $\delta R_{\rm S}=\delta Z+L_{\rm com}/2$, $\hat n=(0,0,1)=\hat Z$ and \eqn{appeq:deltaRS} reduces to the first line in \eqn{appeq:deltaZ-distob}. Further assuming a small box then leads to \eqn{appeq:zob:distob}.

\subsection{Periodicity}
\label{app:pbc}
\noindent
The above did not account for periodic boundary conditions imposed by typical cosmological simulations. For clustering studies which rely on relative distances between multiple tracers, we must also ensure that the periodicity of the simulation box is respected when moving objects into redshift space. This can be done for the general case as follows. 
\begin{itemize}
\item First, use the value of $z_{\rm obs}$ from the cosmic Doppler formula \eqref{appeq:cosmicdoppler-generic} to calculate the new box-centric comoving vector position $\mathbf{x}_{\rm S} = r_{\rm com}(z_{\rm obs})\hat n - \bar Z\,\hat Z$ of the tracer, where $\hat n$ is given by \eqn{appeq:los} and $\bar Z$ by \eqn{appeq:Zbar}.
\item Replace $\mathbf{x}_{\rm S}\to(\mathbf{x}_{\rm S}+\mathbf{b})\%L_{\rm com} - \mathbf{b}$ where $\mathbf{b}\equiv(1,1,1)\times L_{\rm com}/2$, i.e., wrap each coordinate around $L_{\rm com}$ and maintain the centering around the box center.
\item Re-calculate $z_{\rm obs}$ by inverting the relation $r_{\rm com}(z_{\rm obs}) = \left\lVert\mathbf{x}_{\rm S}+\bar Z\,\hat Z\right\rVert$.
\end{itemize}
As a consequence, no observed redshift will correspond to an object outside the comoving space of the box. The scheme above also ensures that no object will have a negative redshift.
For example, in the distant observer limit with $z_{\rm sim}=0$ and a small box, we have $\hat n = \hat Z$ and $r_{\rm com}(z_{\rm obs})\simeq c\,\delta z/H_0 = \delta Z + v_Z/H_0+L_{\rm com}/2$, which must be wrapped around the $Z$-axis of the box.

\emph{Caution:} The scheme above will produce consistent redshift space positions which can be used in clustering studies, but the corresponding \emph{redshifts} themselves do not account for the fact that no two tracers can be more than a comoving distance $\sqrt{3}L_{\rm com}/2$ apart in a periodic box. So the redshifts $z_{\rm obs}$ and redshift space positions $\mathbf{x}_{\rm S}$ should not be combined. The values of $z_{\rm obs}$ would typically be useful in combination with survey selection strategies to assess the impact of selection effects.

\section{Template-Matching}
\label{app:templatematching}

\begin{figure*}
\centering
\includegraphics[width=0.9\textwidth]{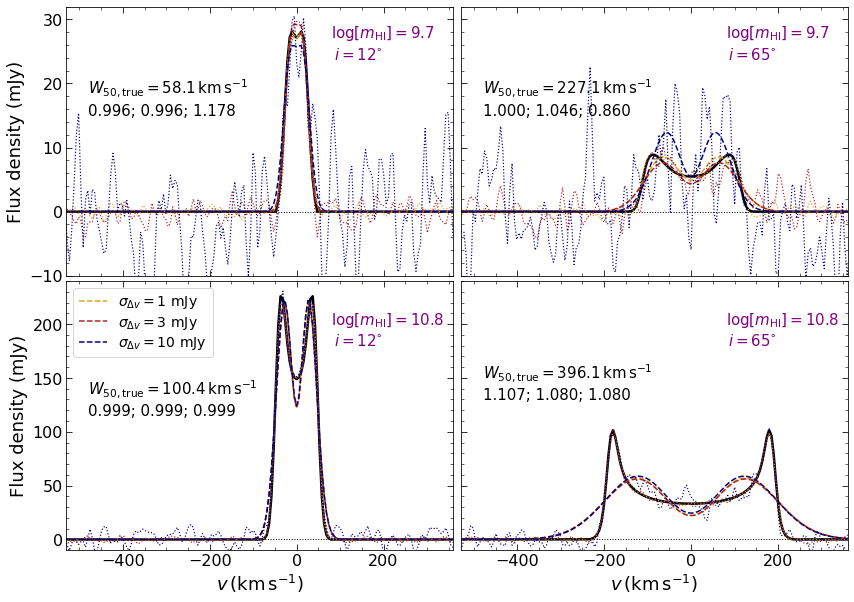}
\caption{{\bf Template-matching performance.} The thick solid black curves in the \emph{top (bottom) row} show the noiseless \Hi\ velocity profile of a realistic mock galaxy with $\log[m_{\Hi}/(\Mhsq)] = 9.7\, (10.8)$, placed at a luminosity distance $D_{\rm L}=100\Mpch$ from the observer and viewed at an inclination of $i=12^\circ$ \emph{(left panels)} and $i=65^\circ$ \emph{(right panels)}.
The true FWHM $W_{50,{\rm true}}$ in each case is indicated as a label. 
Dotted coloured curves show the corresponding noisy profiles with $\sigma_{\Delta v}$ as indicated in the legend of the bottom left panel. 
Dashed coloured curves show the corresponding best-fit template. 
The ratios of the corresponding $W_{50}$ with $W_{50,{\rm true}}$ are displayed as text labels, in order of increasing $\sigma_{\Delta v}$ from left to right. 
Overall, inclination plays a dominant role in causing a systematic offset in the recovered $W_{50}$ ($\sim10\%$ overestimate at low inclinations), with the effects of noise becoming important at low S/N (large $\sigma_{\Delta v}$ coupled with low mass and/or large $D_{\rm L}$; see blue dashed curve in the \emph{top right panel}). 
}
\label{figapp:tempmatch}
\end{figure*}

\noindent
Here we describe a simple algorithm, based on the one presented by \citet{saintonge07}, for performing a robust least-squares estimate of the width $W_{50}$ (or FWHM) of each observed \Hi\ line profile, which is then used for estimating the S/N of the profile, in addition to being an observable in its own right. 

Since the detailed shape of the profile is less relevant at this stage, it is useful to build templates using simple functions with well-defined analytical properties.
Following \citet{saintonge07}, we use the first two symmetric, orthogonal Hermite functions $\Psi_0(v;\sigma)\propto\e{-v^2/(2\sigma^2)}$ and $\Psi_2(v;\sigma)\propto\p_v^2\Psi_0$ (both analytically normalised such that $\int\der v\,|\Psi_n|^2=1$)  to define a template
\be
t(v;\sigma,\lambda) = \Psi_0(v;\sigma) + \lambda\Psi_2(v;\sigma)\,.
\label{appeq:template}
\ee
The signal $s(v)$ is then modelled as $s(v)\sim A\,t(v;\sigma,\lambda)$, with the overall amplitude $A$, width $\sigma$ and relative amplitude $\lambda$ being free parameters.\footnote{Strictly speaking, one should include a fourth parameter $\delta$ to capture the unknown redshift of the galaxy, and model the signal as $s(v)\sim A\,t(v-\delta;\sigma,\lambda)$. For simplicity, we will assume perfect knowledge of each redshift and center all profiles at $\delta=0$.} Realistic signals require $0\leq\lambda\leq\sqrt{2}$, with the lower limit corresponding to the face-on case of a single horn and the upper limit leading to an extreme double horn with zero flux density at $v=0$. 

As discussed by \cite{saintonge07}, a least-squares analysis of the signal $s(v)$ relative to the template $A\,t(v;\sigma,\lambda)$ shows that the best-fitting amplitude $A$ satisfies $A = \sigma_s\,c/\sigma_t$, where $\sigma_s^2$ and $\sigma_t^2$ are the signal and template variance and $c$ is their correlation coefficient. Using this, the $\chi^2$ reduces to $\chi^2\propto\sigma_s^2(1-c^2)$, so that minimising $\chi^2$ is equivalent to maximising $c$. We therefore perform a 2-dimensional maximisation of $c(\sigma,\lambda)$ for a given signal. In detail, we first search for the location of the maximum on a 2-dimensional grid in $(\log\sigma,\lambda)$ and then refine this estimate  using a 5-point interpolation assuming that $c(\log\sigma,\lambda)$ can be approximated by a bi-variate quadratic form in the vicinity of its maximum.
$W_{50}$ is then estimated as the full width of the best-fitting template $t(v;\sigma,\lambda)$ at half of its (common) peak height, with the template being evaluated on the array of velocity channels for the given survey. 

We have checked that this technique accurately recovers the full shape of an injected signal (after adding Gaussian noise), relatively independently of the noise level, when the signal itself is chosen to be one of the templates. 
Turning to the recovery of more realistic signals, Fig.~\ref{figapp:tempmatch} shows the performance of this technique on four injected signals (thick solid black curves) derived from our mock catalog. 
The \emph{top (bottom) row} used a galaxy with $\log[m_{\Hi}/\Mhsq]\simeq9.7\,(10.8)$ placed at $z\simeq0.04$ (luminosity distance $D_{\rm L}=120\Mpch$) and viewed at an inclination $i=12^\circ$ \emph{(left panels)} and $i=65^\circ$ \emph{(right panels)}. 
For each noiseless profile calculated using \eqns{eq:SHI} and~\eqref{eq:mHI-SHI}, we generate three noisy profiles by adding Gaussian noise using values of the per-pixel r.m.s. $\sigma_{\Delta v} = 1,3,10$ mJy, assuming a channel width $\Delta v = 5.3\kms$ (which is appropriate for an ALFALFA-like survey, see appendix~\ref{app:SN}). These examples therefore allow us to explore the effects of inclination  as well as overall S/N on the recovery of $W_{50}$. 

\begin{figure}
\centering
\includegraphics[width=0.47\textwidth,trim=8 10 8 5,clip]{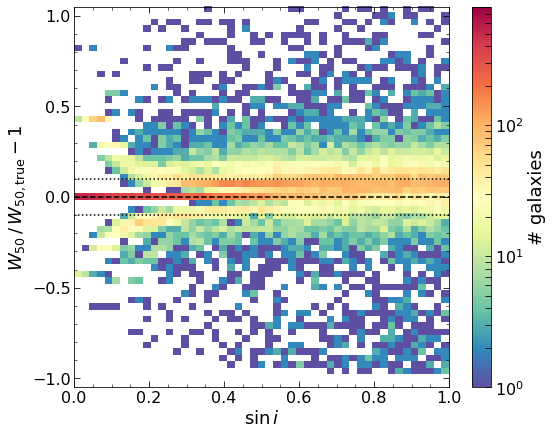}
\caption{{\bf Quality of $W_{50}$ recovery.} Joint distribution of inclination $\sin i$ and relative difference between estimated and true $W_{50}$ for the ALFALFA-like mock sample shown in Fig.~\ref{fig:alfalfa-sampleprops}. We see that low inclinations ($\sin i\lesssim0.4$) lead to essentially perfect recovery while higher inclinations lead to a $\sim10\%$ overestimate of $W_{50}$.}
\label{figapp:W50recovery}
\end{figure}

For most of these cases, it is visually apparent that the best-fitting templates do not exactly match the detailed shape of the input profile, which is not surprising since they are limited by the shapes of the two Hermite functions. Nevertheless, the recovered values of $W_{50}$ differ from the true value $W_{50,{\rm true}}$ by $\lesssim0.5\%$ at low inclination and high S/N. 
We estimate $W_{50,{\rm true}}$ as the full width at half of the (common) peak height of each noiseless profile evaluated on the same discrete velocity channels as the noisy profiles.
Indeed, inclination plays a dominant role in causing a systematic difference between $W_{50}$ and $W_{50,{\rm true}}$, with a $\sim10\%$ overestimate at high inclinations (nearly edge-on galaxies). This is also not surprising, since edge-on disks have sharp peaks in their line profiles separated by a long, flat portion, which cannot be accurately captured by a linear combination of $\Psi_0$ and $\Psi_2$ alone. At large $\sigma_{\Delta v}$, the effect of noise becomes more apparent, especially when combined with a lower signal strength (low $m_{\Hi}$ and/or large $D_{\rm L}$). We now see larger variations in $W_{50}/W_{50,{\rm true}}$ for both high and low inclinations. Fig.~\ref{figapp:W50recovery} shows that, for the ALFALFA-like sample from Fig.~\ref{fig:alfalfa-sampleprops}, the recovery of $W_{50}$ is essentially perfect at $i\lesssim23.5^\circ$, while higher inclinations lead to the $\lesssim10\%$ offset discussed above. 

Overall, these examples show that the template-matching technique described above leads to a reasonably robust recovery ($\sim10\%$ systematic error) of $W_{50}$ for all but the lowest S/N objects. The main text quantifies this further, showing that the mass $m_{\Hi}$ inferred from each profile using its estimated $W_{50}$ deviates substantially from the true mass only at low S/N (see Fig.~\ref{fig:alfalfa-sampleprops}).

\section{Signal-to-noise}
\label{app:SN}
\noindent
As described by \citet{giovanelli+07}, the ALFALFA signal extraction pipeline detailed in \citet{saintonge07} first uses a least-squares template-matching method to produce an initial catalog, with signal-to-noise (S/N) values for each candidate detection determined using the matched templates. A cut is imposed on these S/N values and each object surviving this cut is then visually inspected and processed further. Properties including the velocity width $W_{50}$, integrated flux density $\SHi$ and consequently a S/N ratio depending on these \citep[e.g., equation~16 of][see also below]{saintonge07} are calculated. 

In particular, the integrated flux density is extracted over the `spectral extent' of the signal, which involves a subjective choice for each object \citep[see section~5 of][]{giovanelli+07}.
The initial use of template-matching, as well as the subjective choice of integration range involved in estimating the integrated flux density, leads to a specific relation between $\SHi$ and $W_{50}$ for objects near the threshold of detection, which changes behaviour for $W_{50}\gtrsim400\,\kms$ and is discussed in detail by \citet{giovanelli+07} and \citet{martin+10}.
To simplify our analysis while still keeping it realistic, we do use the template-matching technique described in appendix~\ref{app:templatematching}, but choose to standardise the choice of integration range in estimating the integrated flux density.
We also examine the effects of this standardisation on the statistics of our interest. 

The calculation of S/N requires fixing an integration range of length $\Delta W$ (in  \kms) for the measured velocity profile $S_{\Hi}(v)$, which we assume to be centered on the systemic velocity $c\,z$ of the galaxy. The integrated flux can then be approximated by 
\be
\SHi\simeq\Delta v\sum_{v=-\Delta W/2}^{\Delta W/2}S_{\Hi}(v)\,, 
\label{eq:SHI-int}
\ee
whose measurement error is 
\be
\sigma_{S} = \sqrt{\Delta v\,\Delta W}\,\sigma_{\Delta v}\,.
\label{eq:sigS}
\ee
Similarly to \citet{saintonge07}, we define the S/N as being based on one half of the signal, so that
\be
{\rm S/N} \equiv \frac{\SHi}{\sqrt{2}\,\sigma_S} = \frac{\left(\SHi/\Delta W\right)}{\sigma_{\Delta v}}\left(\frac{\Delta W/2}{\Delta v}\right)^{1/2}\,.
\label{eq:S/N}
\ee
The second equality highlights that the S/N is the ratio of mean flux density over the signal extent to the r.m.s. noise per velocity channel, scaled up by the square-root of the number of independent channels available in half the signal width.
In order to standardise the integration range $\Delta W$ and avoid subjective choices, in the following we will assume 
\be
\Delta W = 1.4\times W_{50}\,,
\label{eq:DW-def}
\ee
with the assumption that the profile will typically contribute only noise in channels with $|v|\gtrsim1.4 (W_{50}/2)$ relative to the central velocity. The value of the prefactor is a compromise between maximising S/N and minimising the bias in the recovery of $m_{\Hi}$; small values of the prefactor will tend to systematically underestimate $m_{\Hi}$, while large  values will integrate over noise and degrade the S/N. We have checked that small variations of the prefactor (values between $1.3$ to $1.6$) do not affect our results. Larger variations (values of, say $1.0$ or $1.8$) lead to a biased inference of the $m_{\Hi}$ and $W_{50}$ abundances relative to the noise-free case, with the bias being relatively insensitive to the chosen S/N threshold. We therefore use \eqn{eq:DW-def} as our default choice in the entire analysis.


The frequency resolution of the ALFALFA observations prior to spectral smoothing is $\Delta\nu = 25\,{\rm kHz}$ \citep{giovanelli+07}. Using $\Delta\nu/\nu_0 = \Delta v/c$ with $\nu_0=1420\,{\rm MHz}\,(1+z)^{-1}$ gives us a channel width 
\be
\Delta  v\simeq 5.3\,\kms\,(1+z)\,. 
\label{eq:Dv-value}
\ee
Spectra are smoothed with a 3-point Hann filter \citep{saintonge07}. This effectively degrades the spectral resolution to $\simeq10\,\kms$ at $z\simeq0$, but does not drastically affect \eqn{eq:S/N} for the S/N, so we will continue to use that  relation in the following. The noise properties of the ALFALFA data cubes \emph{after} Hann smoothing give an r.m.s. $\sigma_{\rm rms}\simeq 2.23\,{\rm mJy}$ \citep[see fig.~4 of][]{saintonge07}, which implies a pre-smoothing value of the per-pixel width $\sigma_{\Delta v}$ of
\be
\sigma_{\Delta v} =  \sqrt{8/3}\,\sigma_{\rm rms} \simeq 3.64\,{\rm mJy}\,,
\label{eq:sigDv-value}
\ee
which we use in our analysis.

\label{lastpage}

\end{document}